\documentclass[journal]{IEEEtran}
\usepackage{cite}

\ifCLASSINFOpdf
   \usepackage[pdftex]{graphicx}
   \graphicspath{{../pdf/}{../jpeg/}}
   \DeclareGraphicsExtensions{.pdf,.jpeg,.png}
\else
   \usepackage[dvips]{graphicx}
   \graphicspath{{../eps/}}
   \DeclareGraphicsExtensions{.eps}
\fi
\usepackage{amsmath}
\usepackage{amssymb}
\usepackage{bm}
\allowdisplaybreaks[3]
\usepackage{color}
\usepackage{pifont}
\usepackage{threeparttable}
\usepackage{booktabs}
\usepackage{multirow}

\newtheorem{proposition}{\bf Proposition}
\newtheorem{remark}{\bf Remark}

\usepackage{array}

\usepackage{url}

\begin{document}
%
\title{Delay Characterization of Mobile Edge Computing for 6G Time-Sensitive Services}
%
%
%

\author{Jianyu~Cao,
        Wei~Feng,~\IEEEmembership{Senior~Member,~IEEE,}
        Ning~Ge,~\IEEEmembership{Member,~IEEE,}
        and~Jianhua~Lu,~\IEEEmembership{Fellow,~IEEE}
\thanks{This work was supported in part by the National Key R\&D Program of China (Grant No. 2018YFA0701601); the National Natural Science Foundation of China (Grant No. 61922049, 61941104, 61771286, 61701457); and the Beijing Innovation Center for Future Chip. \emph{(Corresponding author: Wei Feng.)}}
\thanks{J. Cao, W. Feng, N. Ge and J. Lu are with the Department of Electronic Engineering, Tsinghua University, Beijing 100084, China, and also with the Beijing National Research Center for Information Science and Technology, Tsinghua University, Beijing 100084, China (e-mail: jycao@mail.tsinghua.edu.cn; fengwei@tsinghua.edu.cn; gening@tsinghua.edu.cn; lhh-dee@mail.tsinghua.edu.cn).}}
\maketitle
\begin{abstract}
Time-sensitive services (TSSs) have been widely envisioned for future sixth generation (6G) wireless communication networks. Due to its inherent low-latency advantage, mobile edge computing (MEC) will be an indispensable enabler for TSSs. The random characteristics of the delay experienced by users are key metrics reflecting the quality of service (QoS) of TSSs. Most existing studies on MEC have focused on the average delay. Only a few research efforts have been devoted to other random delay characteristics, such as the delay bound violation probability and the probability distribution of the delay, by decoupling the transmission and computation processes of MEC. However, if these two processes could not be decoupled, the coupling will bring new challenges to analyzing the random delay characteristics. In this paper, an MEC system with a limited computation buffer at the edge server is considered. In this system, the transmission process and computation process form a feedback loop and could not be decoupled. We formulate a discrete-time two-stage tandem queueing system. Then, by using the matrix-geometric method, we obtain the estimation methods for the random delay characteristics, including the probability distribution of the delay, the delay bound violation probability, the average delay and the delay standard deviation. The estimation methods are verified by simulations. The random delay characteristics are analyzed by numerical experiments, which unveil the coupling relationship between the transmission process and computation process for MEC. These results will largely facilitate elaborate allocation of communication and computation resources to improve the QoS of TSSs.
\end{abstract}


%
\IEEEpeerreviewmaketitle

\section{Introduction} \label{sec-1}
With the worldwide deployment of fifth generation (5G) wireless communication networks, increasing research attention has been directed towards sixth generation (6G) wireless communication networks. Although what 6G will be remains open, one certainty is that 6G will support many more time-sensitive services (TSSs) than 5G \cite{LiR-2019}. As a key enabler for low latency guarantees, mobile edge computing (MEC) will dominate 6G to a great extent, leading to a large number of intelligent edge servers for future TSSs.

In \cite{LiR-2019}, both in-time services and on-time services were defined for TSSs. These imply that the delay requirements of 6G will be holistic, including not only the average delay but also other characteristics reflecting random delay fluctuation, e.g., the delay jitter and the delay bound violation probability. The delay jitter denotes the fluctuation of delay over time, which can be measured by the delay standard deviation. For example, in mobile augmented reality/virtual reality (AR/VR) applications, the jitter and delay between consecutive updates would degrade the user experience \cite{FangW-2017}. In \cite{Stauffert-2018}, it was demonstrated that delay jitter would cause cybersickness in head mounted display (HMD)-based VR. In addition, a high motion-to-photon (MTP) delay of 20 ms or more would lead to motion sickness and affect the user visual experience \cite{YangX-2017,Elbamby-2018}.

To effectively support TSSs, we need a holistic delay characterization of MEC. In a mobile AR application, for example, the mobile device offloads the computation tasks, i.e., generating augmented data from raw video streaming, to the edge server. The user experience is closely related to the average delay, delay jitter and delay bound violation probability. All these delay characteristics should be considered in allocating the communication and computation resources\footnote{The communication resources refer to the wireless communication bandwidths allocated to the UE. The computation resources refer to the total number of CPU cycles per second, allocated to the UE.} to the mobile device running the mobile AR application. However, most existing studies on MEC have focused only on the average delay \cite{Lyu-2017,Ye-2020,Wu-2020,WeiZ-2019,LiuL-2018,LiK-2019}, and only a few research efforts have been devoted to other random delay characteristics, such as the delay bound violation probability and the probability distribution of the delay \cite{ZhaoT-2017,WangY-2019}. Different from the existing studies, we consider an MEC system with a limited computation buffer at the edge server. If the computation buffer is saturated, the user equipment (UE) will pause the task transmission until the computation buffer has new free space. This leads to complicated coupling between the task transmission process and computation process when analyzing the random delay characteristics. We formulate a discrete-time two-stage tandem queueing system considering whole-chain delay factors, i.e., the transmission waiting time, transmission time, computation waiting time and computation time. By adopting the matrix-geometric method, the estimation methods for the probability distribution of the delay, the delay bound violation probability, the average delay and the delay standard deviation are obtained. Based on these estimation methods, when an MEC system allocates the communication and computation resources to the user equipment running time-sensitive applications, it can consider various delay characteristics in addition to the average delay. This will effectively improve the quality of service (QoS) of TSSs.

In the following, we first survey the related works on the random delay characteristics of MEC in Section \ref{sec-2} and then present the system model in Section \ref{sec-3}. The estimation methods for the random delay characteristics are given in Section \ref{sec-4} and are followed by numerical evaluations in Section \ref{sec-5}. In Section \ref{sec-6}, some discussions are given. Section \ref{sec-7} concludes this paper and gives the future directions.

%
%

\section{Related work} \label{sec-2}

  At present, the average delay has been widely used in the research on task offloading and communication and computation resource allocation for MEC. For example, the minimum average delay has been directly used as the optimization object in order to provide a higher QoS \cite{LiuJ-2016,Rodrigues-2017}. Liu \MakeLowercase{\textit{et al.}} \cite{LiuJ-2016} formulated a delay minimization problem for an MEC system in which the tasks are executed on a mobile device and an MEC server in parallel. Based on this optimization problem, they developed an efficient one-dimensional search algorithm to find the optimal task scheduling policy. Rodrigues \MakeLowercase{\textit{et al.}} \cite{Rodrigues-2017} proposed a service delay minimization method for an MEC scenario with two edge servers. This method controls the processing delay through virtual machine migration and improves the transmission delay through the transmission power control. Other studies considered the tradeoff between energy consumption and average delay while improving the service delay \cite{ZhangG-2018,MaoY-2017,Meng-2019,MengX-2017,YangY-2018,LiuJ-2019,YangT-2019,XuJ-2017,WangS-2019}. For instance, Zhang \MakeLowercase{\textit{et al.}} \cite{ZhangG-2018} proposed an online dynamic task scheduling method for an MEC system with an energy harvesting capability. This method minimized the average weighted sum of the energy consumption and execution delay of a mobile device while using the stability of buffer queues and the battery level as constraints. Mao \MakeLowercase{\textit{et al.}} \cite{MaoY-2017} proposed an online algorithm for jointly allocating the communication and computational resources in multi-user MEC systems. This algorithm minimized the average weighted sum of the power consumption of mobile devices and an MEC server under the constraint of the average delay. Meng \MakeLowercase{\textit{et al.}} \cite{Meng-2019} proposed a closed-form multi-level water-filling computation offloading scheme for the MEC system with a computation-constrained edge server. This scheme optimizes the average delay based on a Markov decision process framework, and it can achieve better delay performance. Meng \MakeLowercase{\textit{et al.}} \cite{MengX-2017} also proposed a closed-form computation offloading scheme for the hybrid system with cloud servers and fog servers. This scheme has lower energy consumption than the conventional single-type computation offloading under the delay constraint. Due to the limited communication and computation resources, when multiple users concurrently access MEC services, the tradeoff between the profits (or costs) of an MEC system and the average delay must be considered \cite{Pouria-2019,KimY-2018}. Paymard \MakeLowercase{\textit{et al.}} \cite{Pouria-2019} considered a multi-user MEC system with multiple tasks that have different priorities. They proposed a priority-based task scheduling policy to maximize the profits of a mobile network operator under the constraint of the average delay. Kim \MakeLowercase{\textit{et al.}} \cite{KimY-2018} explored the cost-delay tradeoffs for mobile users and a code offloading service provider (CSP) in an MEC system. They proposed three different algorithms to minimize the energy/monetary costs while ensuring a finite processing delay under a competitive scenario. In addition, the service fairness in terms of the average delay is also considered \cite{GaoH-2019}. The fairness index reaches its maximum when all users experience the same delay \cite{YinH-2017}.

  In contrast to the average delay that has been widely considered, a few research efforts have studied other random delay characteristics. Zhao \MakeLowercase{\textit{et al.}} \cite{ZhaoT-2017} considered an MEC scenario in which the edge server and remote cloud server serve multiple users. They jointly optimized the task scheduling among heterogeneous servers and the computation resource allocation in the edge server for multiple users by maximizing the probability that the task delay does not exceed the required bound. Wang \MakeLowercase{\textit{et al.}} \cite{WangY-2019} considered an MEC scenario in which the edge server serves multiple users. They formulated the problem, i.e., the task offloading decision and communication and computation resource allocation, to maximize the total revenues of the MEC system while guaranteeing that the delay bound violation probability does not exceed the given bound. She \MakeLowercase{\textit{et al.}} \cite{SheC-2019} studied the packet offloading and communication resource allocation scheme for mission-critical Internet-of-Things (MC-IoT) services with short packets. They derived the probability distribution of the delay experienced by short packets and minimized the overall packet loss probability caused by a delay bound violation.

  The above studies on MEC addressed the delay bound violation probability or the probability distribution of the delay by decoupling the transmission process and computation process. However, if these two processes form a feedback loop in some MEC scenarios, they could not be decoupled. This coupling will bring new challenges to analyzing the random delay characteristics, especially the probability distribution of the delay. Different from the existing studies, we will consider an MEC system with a limited computation buffer at the edge server. If the computation buffer is saturated, the user equipment will pause the task transmission until the computation buffer has new free space. We formulate a discrete-time two-stage tandem queueing system. By jointly considering the transmission process and computation process, the probability distribution of the delay, the delay bound violation probability, the average delay and the delay standard deviation will be derived.

  The delay in MEC has been analyzed by modeling the MEC system as a tandem queueing system. Guo \MakeLowercase{\textit{et al.}} \cite{GuoS-2018} proposed a tandem queueing system for a base station that serves multi-class uplink and downlink transmission and MEC users simultaneously. Then, they derived the average delay of each user class of each service type by decomposing the tandem queueing system into three single-queue systems, based on the Burke theorem. Chang \MakeLowercase{\textit{et al.}} \cite{ChangP-2018} considered an MEC system consisting of multiple base stations and one MEC server. They formulated the whole system as several two-stage tandem queueing systems and studied the trade-off between the energy consumption and the service latency by decomposing the tandem queueing systems based on the Burke theorem. The Burke theorem-based decomposition method is available, if the transmission process is independent of the computation process and the inter-arrival time and transmission time of tasks follow the exponential distribution. Compared with \cite{GuoS-2018,ChangP-2018}, the Burke theorem-based decomposition method is not available in this paper. The matrix-geometric method will be used to analyze the random delay characteristics by jointly considering the queues of the tandem queueing system.

\section{System model} \label{sec-3}

\begin{figure}[!t]
\centering
\includegraphics[width=3.5in]{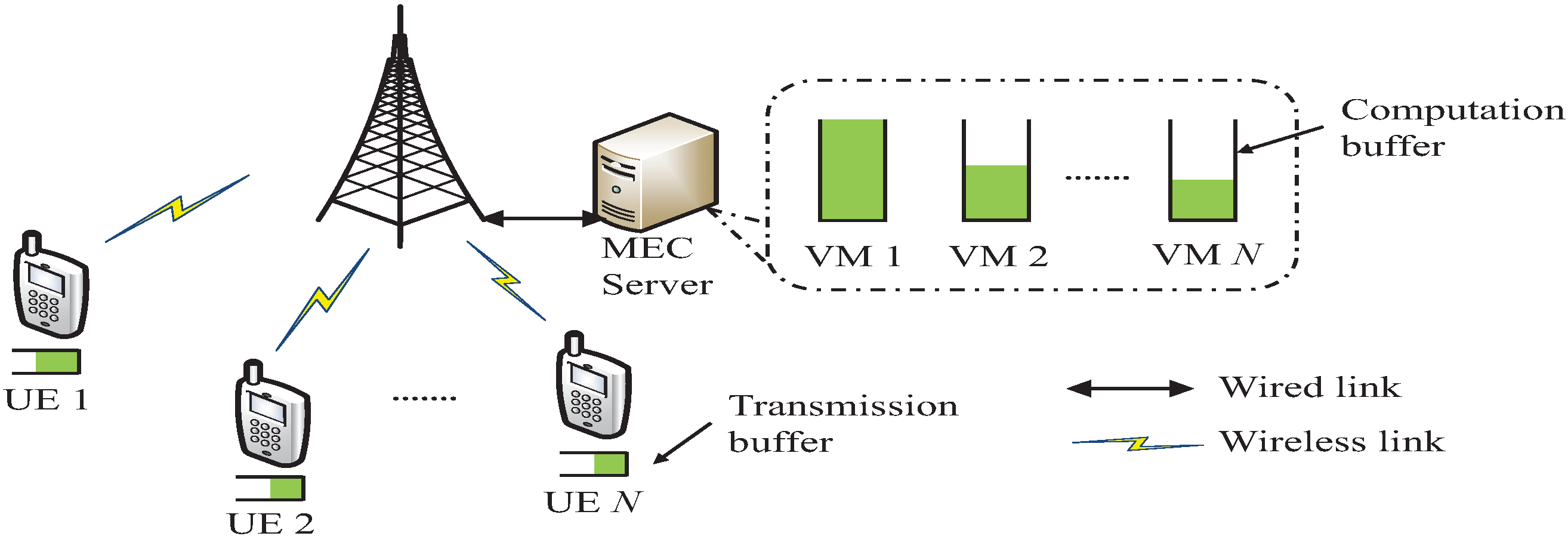}
\caption{The task offloading process in an MEC scenario.}
\label{fig-01}
\end{figure}

Computation offloading is preferred to local computation for UEs with limited energy and desirable channel condition \cite{Mao-2017}, especially for battery-powered IoT devices that have limited computation capability but are expected to perform sophisticated tasks such as smart IoT sensing, positioning and tracking \cite{WangB-2018} and VR/AR. Accordingly, a typical MEC scenario, shown in Fig. \ref{fig-01}, will be considered in this paper.

An edge server is deployed in the mobile access network, and it assigns a virtual machine (VM), which is computationally constrained and has a limited computation buffer, to each UE requesting the computing service. It is stated in the MEC specification of the European Telecommunications Standards Institute (ETSI) that one application instance per user is supported in the MEC server \cite{ETSI-MEC-2016}. And the application instance is created in an individual VM \cite{ETSI-MEC-2019}. Thus, the ``user-to-VM'' association \cite{Ceselli-2018} is typical in MEC. For instance, a similar system model with ``user-to-VM'' was considered in \cite{ZhaoT-2017}. The ``user-to-VM'' association is also typical for IoT, since the users in the MEC specification of ETSI are general and include IoT devices \cite{ETSI-MECW-2019}. Suppose that after a task is generated in the UE, if the free space of the transmission buffer cannot accommodate the task, the task will be computed locally; otherwise, the task enters the transmission buffer. The tasks entering the transmission buffer will be offloaded to the corresponding VM in first-in-first-out (FIFO) order. It is assumed that the wireless communication network has high reliability. Since the propagation distance from the UE to the edge server is short, the tasks can be successfully received by the edge server in the same order as they were transmitted. If the computation buffer has free space, then the tasks in the receiving buffer of the transfer control protocol stack can be read to the computation buffer immediately. In this paper, the receiving buffer and computation buffer are collectively referred to as the computation buffer. According to the traffic control mechanism commonly used in the task transmission process, when the computation buffer is saturated, the UE will pause the task transmission until the computation buffer has free space. For instance, in the transmission control protocol (TCP), the sender can adjust the data transmission rate according to the acknowledgement returned by the receiver, since the acknowledgement contains the size of the receiving buffer. Thus, the synchronization of sending rate and receiving ability can be realized. In the Quick UDP Internet Connection (QUIC) protocol proposed by Google, a similar acknowledgement is also used to synch the sending rate and receiving ability. Suppose that in each VM, the tasks are queued in the computation buffer and computed in FIFO order.

In the above MEC system, the UE gives priority to offloading the tasks to the edge, and the local computation is selected only when the transmission buffer is saturated. However, if the task arrival rate is high, the local processor of the IoT device with limited energy and computation capability may not satisfy the delay requirements of all the tasks that cannot enter the transmission buffer. When this occurs, it is necessary to allocate more communication and computation resources to increase the offloading ratio, i.e., more tasks can be offloaded to the edge server immediately. This issue is out of the scope of this paper, but it indicates an interesting future work direction on the joint optimization of communication and computation resources under the constraints of task execution delay and device energy consumption.

\begin{figure}[!t]
\centering
\includegraphics[width=3.4in]{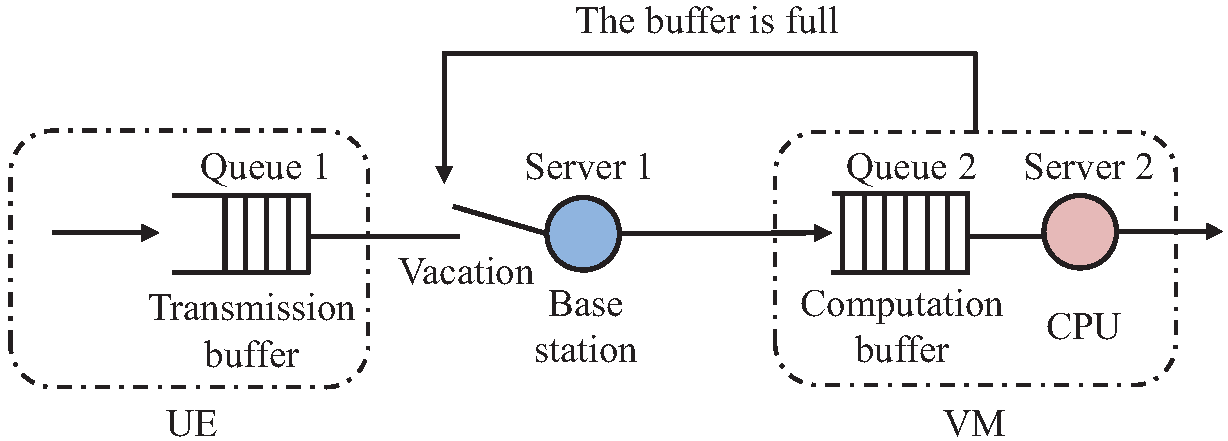}
\caption{A diagram of the two-stage tandem queueing system.}
\label{fig-02}
\end{figure}

For the above MEC scenario, we mainly study the estimation methods for the random delay characteristics, including the probability distribution of the delay, the delay bound violation probability, the average delay and the delay standard deviation. We assume that the data size of the computation result is small relative to the offloaded data, so that the time during which the computation result is returned to the UE from the edge server is ignored. The delay considered in this paper is caused by two processes. The first is the task transmission process, which lasts from the instant when the task enters the transmission buffer to the instant when the task is transmitted to the edge server. The second is the task computation process, which lasts from the instant when the task arrives at the edge server to the instant when the computation is completed in the VM. In other words, the delay consists of the transmission waiting time, the transmission time, the computation waiting time and the computation time. To obtain the estimation methods for random delay characteristics, the discrete-time two-stage tandem queueing system shown in Fig. \ref{fig-02} is constructed. The time is slotted with equal length $\Delta t$ $(0< \Delta t<\infty )$, and the time slots are indexed by $t_i$, where $t_0=0$ and $t_{i+1}=t_{i}+\Delta t$ $(i=0,1,2,\cdots)$. Based on the tandem queueing system shown in Fig. \ref{fig-02}, we establish the delay estimation methods for the single-user case. Further, given multi-access strategies, these methods can be extended to the multi-user case. This, however, is not within the scope of this work because if the transmission and computation processes could not be decoupled, then the estimation methods for holistic random delay characteristics of the benchmark single-user case remain open.\footnote{Currently, if the transmission and computation processes could not be decoupled, the existing works focused on the average delay \cite{Meng-2019}. The estimation methods for holistic random delay characteristics have been derived for the single-user case in which the transmission and computation processes could be decoupled, and they were applied to the resource allocation for the multi-user case under the orthogonal access strategy \cite{ZhaoT-2017,WangY-2019}.} It will be important to consider the multi-user case in future works. The descriptions for the main elements of the tandem queueing systems are as follows.

  For the task arrival process, assume that the tasks have the same size in bits and that they are generated according to the discrete-time Markovian arrival process (D-MAP). The D-MAP can capture the correlated and bursty nature of the task generation process. Moreover, the D-MAP includes many familiar arrival processes, such as the Bernoulli arrival process, the discrete-time phase type renewal (PH-renewal) process, and the Markov-modulated Bernoulli process (MMBP) \cite{Gupta-2007}. For more details on D-MAP, refer to Appendix \ref{Appendix-A}.

  Queue 1 represents the transmission buffer of the UE and has limited space. The tasks arrive at queue 1 in order of their generation. Upon arrival, if there is not enough free space in queue 1, the task will be rejected; otherwise, the task enters queue 1. The rejected task is computed locally. The local computation process is not considered in this paper.

  Queue 2 represents the computation buffer of the VM and has limited space. The tasks enter queue 2 after departing from queue 1.

  Server 1 represents the base station. For the convenience of analysis, the process, during which the tasks are transmitted to the base station and then forwarded to the edge server successfully, is abstracted as the process during which server 1 renders transmission services to the tasks in queue 1. The wireless links are not error-free, and the channel condition is random. The distribution followed by the transmission time of each task depends on the channel condition, the power of the transmitter, the allocated bandwidth, the size of the task, etc. It is difficult to strictly obtain the transmission time distribution. To simplify the delay analysis of an MEC system based on queueing theory, the transmission time of each task is assumed to obey the exponential distribution in \cite{GuoS-2018,ChangP-2018,Alnoman-2019}. In this paper, it is assumed that the service time for each task obeys a discrete-time phase type (D-PH) distribution. Because the PH distribution is general and can approximate any probability distribution on $[0,+\infty)$ \cite{BreuerBaum-2005} by setting appropriate parameters.\footnote{Some software packages for parameter estimation of PH distributions are available online: http://webspn.hit.bme.hu/{\%7e}telek/tools.htm.} There are two types of PH distributions: continuous PH (C-PH) distribution and D-PH distribution. Through discretization, the continuous distribution can be approximated by the discrete-time distribution. For example, the Weibull distribution can be accurately approximated using the D-PH distribution, as shown in \cite{Isensee-2005}. The widely-used Rayleigh distribution is a special case of the Weibull distribution, and it may also be approximated using the D-PH distribution. For more details on PH distribution, refer to Appendix \ref{Appendix-B}.

  Server 2 represents the CPU of the VM. Assume that the computation time needed by each task obeys a D-PH distribution.

  The service discipline attached to queue 1 is as follows. The tasks are served by server 1 according to FIFO order. After a task is transmitted to queue 2, if queue 1 is empty, server 1 waits until a new task arrives. If queue 1 is not empty and queue 2 is not full, server 1 continues to serve the tasks in queue 1; otherwise, if queue 1 is not empty and queue 2 is full, server 1 goes on vacation. After the vacation expires, if queue 2 is not full, server 1 immediately renders transmission services to the tasks in queue 1; otherwise, server 1 goes on vacation again. Assume that one vacation duration obeys a D-PH distribution. In the TCP, if the acknowledgement returned by the receiver indicates that the receiving buffer is full, the sender will pause the task transmission for a period of time. After this period expires, the sender asks the receiver if there is new free space. If there is new free space, the sender restarts the transmission immediately; otherwise, it continues to pause the task transmission for a period of time. Thus, one vacation of server 1 represents one pause at the sender.

  The service discipline attached to queue 2 is as follows. Server 2 uses the exhaustive service discipline to render computing services to the tasks in queue 2 according to FIFO order.

  The two-stage tandem queueing system has the following parameters. The buffer sizes of queue 1 and queue 2 are ${{N}_{1}}$ and ${{N}_{2}}$ tasks, respectively, where $1\le {{N}_{1}}<{{N}_{2}}<\infty $. In practical applications, the size of transmission buffer in the UE (especially, the battery-powered IoT device) is usually smaller than the size of computation buffer in the VM allocated to the UE. Moreover, the analysis methods for cases with different relationships between $N_1$ and $N_2$ are the same. Hence, we mainly analyze the case with $N_1 < N_2$ in this paper. The task arrival process is a D-MAP whose parameters are $m$-dimensional matrices ${{D}_{0}}$ and ${{D}_{1}}$ $(1\le m<\infty )$, and let $D=D_0+D_1$. The average arrival rate is denoted by $\lambda $ $\left( 0<\lambda <\infty  \right)$. The transmission times of the tasks in queue 1 are independent and identically distributed random variables, which obey a D-PH distribution with the representation $({{\beta }_{1}},{{S}_{1}})$ of order ${{n}_{1}}\left( 1\le {{n}_{1}}<\infty  \right)$ and the mean ${{b}_{1}}$ $\left( \Delta t<{{b}_{1}}<\infty  \right)$. The average transmission rate is denoted by $\mu_1=1/b_1$. The computation times of the tasks in queue 2 are independent and identically distributed random variables, which obey a D-PH distribution with the representation $({{\beta }_{2}},{{S}_{2}})$ of order ${{n}_{2}}\left( 1\le {{n}_{2}}<\infty  \right)$ and the mean ${{b}_{2}}$ $\left( \Delta t<{{b}_{2}}<\infty  \right)$. The average computation rate is denoted by $\mu_2=1/b_2$. The vacation duration of server 1 obeys a D-PH distribution with the representation $(v,V)$ of order ${{l}_{2}}\left( 1\le {{l}_{2}}<\infty  \right)$ and the mean ${{\eta }_{2}}$ $\left( \Delta t<{{\eta }_{2}}<\infty  \right)$.

  In this paper, for the convenience of analysis, some notations are used as follows. We define $\mathbf{0}$ as a vector or matrix of appropriate size consisting of 0's, define ${{\mathbf{0}}_{{{k}_{1}}\times {{k}_{2}}}}$ as a ${{k}_{1}}\times {{k}_{2}}$ matrix consisting of 0's, where ${{k}_{1}},{{k}_{2}}\in\{1,2,3,\cdots\}$, define $\mathbf{e}$ as a column vector of appropriate size consisting of 1's, define ${{\mathbf{1}}_{{{k}_{1}}\times {{k}_{2}}}}$ as a ${{k}_{1}}\times {{k}_{2}}$ matrix consisting of 1's, define $\mathbf{I}$ as an identity matrix of appropriate size, and define $P\{A\}$ as the probability that event $A$ occurs.

  In the MEC scenario considered in this paper, the transmission process is related to the state of the computation buffer, and so the transmission process and computation process could not be decoupled when analyzing the random delay characteristics. Unfortunately, the existing estimation methods for the random delay characteristics of MEC, especially the probability distribution of the delay, the delay bound violation probability and the delay standard deviation, are not applicable. For this problem, we construct the above discrete-time two-stage tandem queueing system. To the best of our knowledge, this tandem queueing system has not been studied regarding the random delay characteristics that concern us. The works in \cite{Alfa-1995,Baumann-2017} are closely related to our tandem queueing system. Alfa \cite{Alfa-1995} considered a vacation queueing model with single queue, single server and an infinite buffer, in discrete time. He analyzed the probability distributions of the number of customers in the system at arbitrary time and of the waiting time. Compared with the model in \cite{Alfa-1995}, the main differences of our model are that, ours is a tandem queueing model consisting of two single-queue single-server queueing systems each of which has a finite buffer; moreover, the first queueing system is a vacation one in which the trigger condition of vacation relies on the state of the second queue. Baumann \MakeLowercase{\textit{et al.}} \cite{Baumann-2017} considered a tandem queueing model consisting of two single-queue multi-server queueing systems each of which has a finite buffer, in continuous time. They analyzed the impact of various parameters on the loss probability, the blocking probability, and the expected numbers of customers in the two queues. Compared with the model in \cite{Baumann-2017}, the main differences of our model are that, the first queueing system is a vacation one in which the trigger condition of vacation relies on the state of the second queue; moreover, the customer arrival process is the D-MAP, and the service time distribution of each customer in two queues is the D-PH distribution.

  \begin{remark}
    In the theoretical analysis for queueing systems, the main differences between finite and infinite buffer assumptions are as follows. If the buffer of a queueing system is infinite, the queue length may increase to infinity, which makes the queueing system unstable. Thus, the stability must be considered. In addition, since the buffer always has enough free space, it is not needed to consider the effect of blocking on the system state transitions. In contrast, under the finite buffer assumption, the stability disappears and is replaced by blocking. In this paper, when the buffer of queue 2 is full, the tasks in queue 1 are blocked. This complicates the analysis of system state transitions. Since the infinite buffer assumption cannot present the blocking characteristic of the queueing system with the finite buffer, it is infeasible to derive the delay characteristics through assuming that the buffer is infinite first and then amending the results. However, it is feasible to derive the delay characteristics of the queueing system with the infinite buffer by assuming the buffer is finite first and then gradually increasing the buffer size.
  \end{remark}

\section{Estimation methods for random delay characteristics} \label{sec-4}

By considering the states $\xi(t_i)$ of the tandem queueing system at time ${{t}_{i}}$ $(i=0,1,2,\cdots)$, a homogeneous Markov chain $\Xi=\{\xi(t_i);i=0,1,2,\cdots\}$ is constructed on the state space ${\Psi }$ as
\begin{equation}
{\Psi }=\left\{ \left( {{i}_{1}},{{i}_{2}},\bm{\varphi } \right):0\le {{i}_{1}}\le {{N}_{1}},0\le {{i}_{2}}\le {{N}_{2}},\bm{\varphi }\in {{\Phi }_{{{i}_{1}},{{i}_{2}}}} \right\},
\end{equation}
i.e., $\xi(t_i)\in{\Psi }$. Each state consists of two parts, i.e., the level and the phase. Both ${{i}_{1}}$ and ${{i}_{2}}$ are called the level, and they represent that there are ${{i}_{1}}$ and ${{i}_{2}}$ tasks in queue 1 and queue 2, respectively. The task being served is also counted in the queue length. The component $\bm{\varphi }$ denotes the phase belonging to the phase set ${{\Phi }_{{{i}_{1}},{{i}_{2}}}}$ as follows:
\begin{align}
  & {{\Phi }_{0,0}}=\mathsf{\mathbb{M}};\\
  & {{\Phi }_{0,{{i}_{2}}}}=\mathsf{\mathbb{M}}\times {{\mathsf{\mathbb{N}}}_{2}}, \quad\quad\quad\quad\quad\;\, 1\le {{i}_{2}}\le {{N}_{2}};\\
  & {{\Phi }_{{{i}_{1}},0}}=\mathsf{\mathbb{M}}\times {{\mathsf{\mathbb{L}}}_{2}}\times\{v\}\cup \mathsf{\mathbb{M}}\times {{\mathsf{\mathbb{N}}}_{1}}\times\{s\},\nonumber\\
  & \phantom{{{\Phi }_{{{i}_{1}},0}}=\mathsf{\mathbb{M}}\times {{\mathsf{\mathbb{L}}}_{2}}\times\{v\}} \quad\quad\quad\; 1\le {{i}_{1}}\le {{N}_{1}};\\
  & {{\Phi }_{{{i}_{1}},{{i}_{2}}}}=\mathsf{\mathbb{M}}\times {{\mathsf{\mathbb{L}}}_{2}}\times {{\mathsf{\mathbb{N}}}_{2}}\times\{v\}\cup \mathsf{\mathbb{M}}\times {{\mathsf{\mathbb{N}}}_{1}}\times {{\mathsf{\mathbb{N}}}_{2}}\times\{s\}, \nonumber\\
  & \phantom{{{\Phi }_{0,{{i}_{2}}}}=\mathsf{\mathbb{M}}\times {{\mathsf{\mathbb{L}}}_{2}}\times {{\mathsf{\mathbb{N}}}_{2}}}1\le {{i}_{1}}\le {{N}_{1}},\,1\le {{i}_{2}}<{{N}_{2}};\\
  & {{\Phi }_{{{i}_{1}},{{N}_{2}}}}=\mathsf{\mathbb{M}}\times {{\mathsf{\mathbb{L}}}_{2}}\times {{\mathsf{\mathbb{N}}}_{2}}, \quad\quad\;\; 1\le {{i}_{1}}\le {{N}_{1}}.
\end{align}
$\mathsf{\mathbb{M}}=\left\{ 1,2,\cdots ,m \right\}$, ${{\mathsf{\mathbb{L}}}_{2}}=\left\{ 1,2,\cdots,{{l}_{2}} \right\}$ and ${{\mathsf{\mathbb{N}}}_{i}}=\left\{ 1,2,\cdots ,{{n}_{i}} \right\}$ for $i=\{1,2\}$. The definition of the phase $\bm{\varphi }$ is given in Table \ref{Tab-0}.
\begin{table*}[!t]
\centering
\caption{The definitions of the phases in the states of Markov chain $\Xi$}
\label{Tab-0}
\renewcommand\arraystretch{1.5}
    \begin{threeparttable}
    \begin{tabular}{p{0.15\textwidth} p{0.75\textwidth}}
        \hline
        Phase & Definition{\tnote{1}}\\
        \hline
        $\bm{\varphi }\in{{\Phi }_{0,0}}$ & $\bm{\varphi }=\varsigma$ represents that the phase of the D-MAP is $\varsigma$. \\
        \specialrule{0em}{0.75pt}{0.75pt}
        $\bm{\varphi }\in{{\Phi }_{0,i_2}}$ & $\bm{\varphi }=(\varsigma,j_2)$ represents that the phase of the D-MAP is $\varsigma$ and the phase of the D-PH distribution $(\beta_{2},S_{2})$ is $j_2$.   \\
        \specialrule{0em}{0.75pt}{0.75pt}
        $\bm{\varphi }\in{{\Phi }_{i_1,0}}$ & $\bm{\varphi }=(\varsigma,j,v)$ represents that the phase of the D-MAP is $\varsigma$ and the phase of the D-PH distribution $(v,V)$ is $j$;\\
        &$\bm{\varphi }=(\varsigma,j_1,s)$ represents that the phase of the D-MAP is $\varsigma$ and the phase of the D-PH distribution $(\beta_{1},S_{1})$ is $j_1$.\\
        \specialrule{0em}{0.75pt}{0.75pt}
        $\bm{\varphi }\in{{\Phi }_{i_1,i_2}}$ & $\bm{\varphi }=(\varsigma,j,j_2,v)$ represents that the phase of the D-MAP is $\varsigma$, the phase of the D-PH distribution $(v,V)$ is $j$, and the phase of the D-PH distribution $(\beta_{2},S_{2})$ is $j_2$;\\
        &$\bm{\varphi }=(\varsigma,j_1,j_2,s)$ represents that the phase of the D-MAP is $\varsigma$, the phase of the D-PH distribution $(\beta_{1},S_{1})$ is $j_1$, and the phase of the D-PH distribution $(\beta_{2},S_{2})$ is $j_2$.\\
        $\bm{\varphi }\in{{\Phi }_{{{i}_{1}},{{N}_{2}}}}$ & $\bm{\varphi }=(\varsigma,j,j_2)$ represents that the phase of the D-MAP is $\varsigma$, the phase of the D-PH distribution $(v,V)$ is $j$, and the phase of the D-PH distribution $(\beta_{2},S_{2})$ is $j_2$.\\
        \hline
        \end{tabular}
        \begin{tablenotes}
            \footnotesize
            \item[1] $\varsigma\in\mathsf{\mathbb{M}}$, $j\in{\mathsf{\mathbb{L}}}_{2}$, $j_1\in{{\mathsf{\mathbb{N}}}_{1}}$ and $j_2\in{{\mathsf{\mathbb{N}}}_{2}}$. The tasks are generated according to the D-MAP; the transmission time of each task in queue 1 obeys the D-PH distribution $(\beta_{1},S_{1})$; the transmission time of each task in queue 2 obeys the D-PH distribution $(\beta_{2},S_{2})$; the vacation duration of server 1 obeys the D-PH distribution $(v,V)$.
        \end{tablenotes}
    \end{threeparttable}
\end{table*}
\begin{figure*}[!t]
\centering
\includegraphics[width=6.5in]{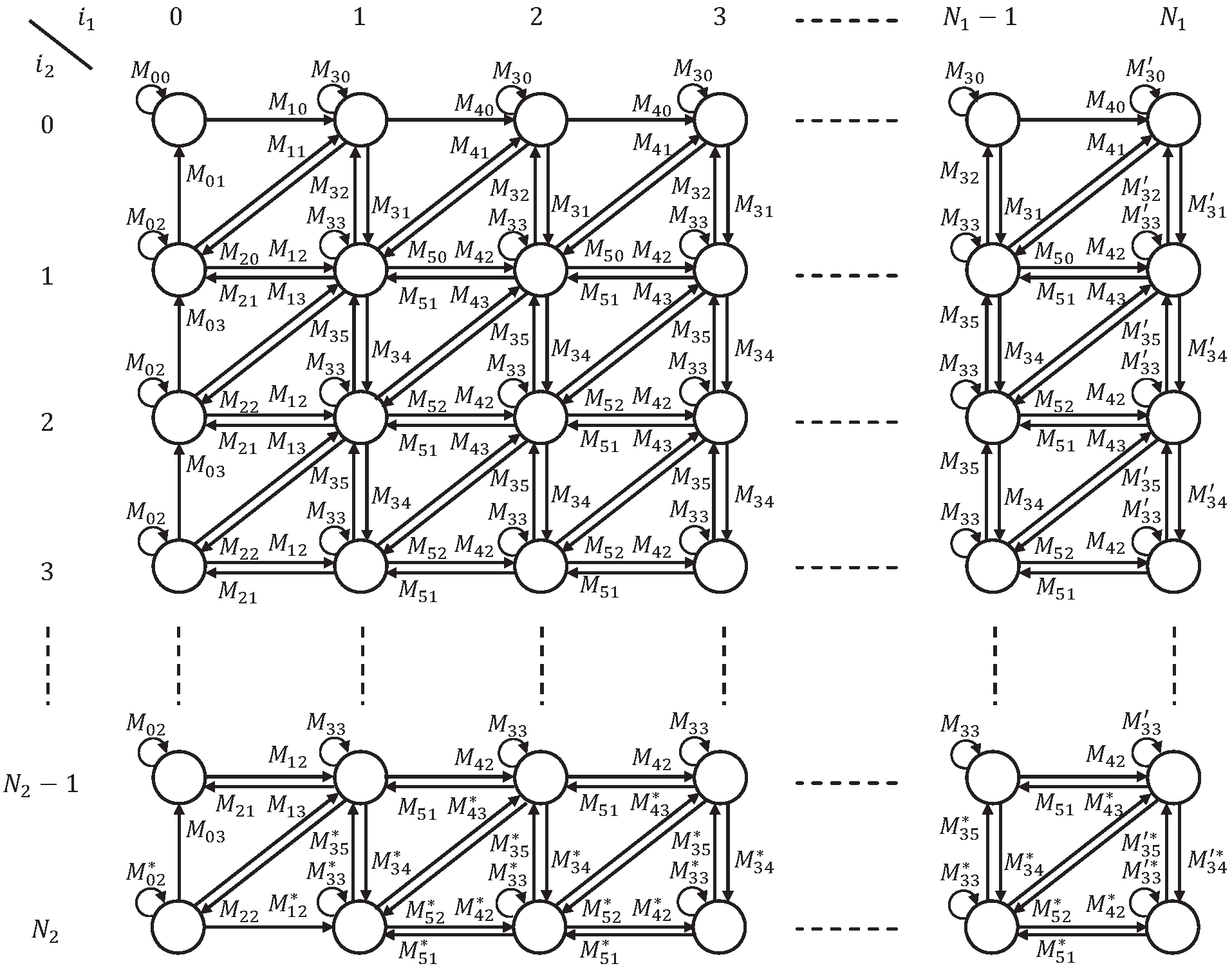}
\caption{A diagram of the one-step transitions among the states in $\Psi $.}
\label{fig-03}
\end{figure*}

The one-step transitions among the states in $\Psi$ are shown in Fig.\ref{fig-03}. The circle for column $i_1$ and row $i_2$ represents the state subspace ${{\Psi }_{{{i}_{1}},{{i}_{2}}}}=\left\{ \left( {{i}_{1}},{{i}_{2}},\bm{\varphi } \right):\bm{\varphi }\in {{\Phi }_{{{i}_{1}},{{i}_{2}}}} \right\}$ whose states are arranged in the order of their lexicography. The matrix, which is beside the arrow from the circle with index $(i_1,i_2)$ to the circle with index $(i'_1,i'_2)$, denotes the one-step transition probability matrix from the states in ${{\Psi }_{{{i}_{1}},{{i}_{2}}}}$ to the states in ${\Psi }_{{i'_1},{i'_2}}$. For example, the matrix $M_{35}$, which is beside the arrow from the circle with index $(1,3)$ to the cycle with index $(1,2)$, denotes the one-step transition probability matrix from the states in $\{(1,3,\bm{\varphi}):\bm{\varphi}\in{{\Phi }_{1,3}}\}$ to the states in $\{(1,2,\bm{\varphi}):\bm{\varphi}\in{{\Phi }_{1,2}}\}$. In addition, $M_{35}$ also denotes the one-step transition probability matrix from the phases in ${{\Phi }_{1,3}}$ to the phases in ${{\Phi }_{1,2}}$. The one-step transition probability matrix of the Markov chain $\Xi$ is constructed as the $(N_1+1)\times(N_1+1)$ block matrix $P$:
\begin{equation}
  P=\left( \begin{matrix}
   {{M}_{0}} & {{M}_{1}} & \mathbf{0} & \mathbf{0} & \cdots  & \mathbf{0} & \mathbf{0}  \\
   {{M}_{2}} & {{M}_{3}} & {{M}_{4}} & \mathbf{0} & \cdots  & \mathbf{0} & \mathbf{0}  \\
   \mathbf{0} & {{M}_{5}} & {{M}_{3}} & {{M}_{4}} & \cdots  & \mathbf{0} & \mathbf{0}  \\
   \mathbf{0} & \mathbf{0} & {{M}_{5}} & {{M}_{3}} & \cdots  & \mathbf{0} & \mathbf{0}  \\
   \vdots  & \vdots  & \vdots  & \vdots  & \ddots  & \vdots  & \mathbf{0}  \\
   \mathbf{0} & \mathbf{0} & \mathbf{0} & \mathbf{0} & \cdots  & {{M}_{3}} & {{M}_{4}}  \\
   \mathbf{0} & \mathbf{0} & \mathbf{0} & \mathbf{0} & \cdots  & {{M}_{5}} & {{M}'_{3}}  \\
  \end{matrix} \right),
\end{equation}
where the matrix blocks are all $(N_2+1)\times(N_2+1)$ block matrices and have the following respective structures. More detailed expressions of these matrix blocks are given in Appendix \ref{Appendix-D}.

      \begin{equation}
           {{M}_{0}}=\left( \begin{matrix}
           {{M}_{00}} & \mathbf{0} & \mathbf{0} & \cdots  & \mathbf{0} & \mathbf{0}  \\
           {{M}_{01}} & {{M}_{02}} & \mathbf{0} & \cdots  & \mathbf{0} & \mathbf{0}  \\
           0 & {{M}_{03}} & {{M}_{02}} & \cdots  & \mathbf{0} & \mathbf{0}  \\
           \vdots  & \vdots  & \vdots  & \ddots  & \vdots  & \vdots   \\
           \mathbf{0} & \mathbf{0} & \mathbf{0} & \cdots  & {{M}_{02}} & \mathbf{0}  \\
           \mathbf{0} & \mathbf{0} & \mathbf{0} & \cdots  & {{M}_{03}} & M_{02}^{*}  \\
          \end{matrix} \right).
      \end{equation}

      \begin{equation}
          {{M}_{1}}=\left( \begin{matrix}
           {{M}_{10}} & \mathbf{0} & \mathbf{0} & \cdots  & \mathbf{0} & \mathbf{0}  \\
           {{M}_{11}} & {{M}_{12}} & \mathbf{0} & \cdots  & \mathbf{0} & \mathbf{0}  \\
           \mathbf{0} & {{M}_{13}} & {{M}_{12}} & \cdots  & \mathbf{0} & \mathbf{0}  \\
           \vdots  & \vdots  & \vdots  & \ddots  & \vdots  & \vdots   \\
           \mathbf{0} & \mathbf{0} & \mathbf{0} & \cdots  & {{M}_{12}} & \mathbf{0}  \\
           \mathbf{0} & \mathbf{0} & \mathbf{0} & \cdots  & {{M}_{13}} & M_{12}^{*}  \\
          \end{matrix} \right).
      \end{equation}

      \begin{equation}
        {{M}_{2}}=\left( \begin{matrix}
           {{\mathbf{0}}_{\tau_1\times m}} & {{M}_{20}} & \mathbf{0} & \cdots  & \mathbf{0} & \mathbf{0}  \\
           \mathbf{0} & {{M}_{21}} & {{M}_{22}} & \cdots  & \mathbf{0} & \mathbf{0}  \\
           \mathbf{0} & \mathbf{0} & {{M}_{21}} & \cdots  & \mathbf{0} & \mathbf{0}  \\
           \vdots  & \vdots  & \vdots  & \ddots  & \vdots  & \vdots   \\
           \mathbf{0} & \mathbf{0} & \mathbf{0} & \cdots  & {{M}_{21}} & {{M}_{22}}  \\
           \mathbf{0} & \mathbf{0} & \mathbf{0} & \cdots  & \mathbf{0} & {{\mathbf{0}}_{\tau_2\times\tau_2}}  \\
        \end{matrix} \right),
      \end{equation}
      where $\tau_1= m\cdot {{l}_{2}}+m\cdot {{n}_{1}}$ and $\tau_2 = m\cdot {{l}_{2}}\cdot {{n}_{2}}$.

      \begin{equation}
        {{M}_{3}}=\begin{pmatrix}
        {{M}_{30}} & {{M}_{31}} & \mathbf{0} & \cdots  & \mathbf{0} & \mathbf{0}  \\
        {{M}_{32}} & {{M}_{33}} & {{M}_{34}} & \cdots  & \mathbf{0} & \mathbf{0}  \\
        \mathbf{0} & {{M}_{35}} & {{M}_{33}} & \cdots  & \mathbf{0} & \mathbf{0}  \\
        \vdots  & \vdots  & \vdots  & \ddots  & \vdots  & \vdots   \\
        \mathbf{0} & \mathbf{0} & \mathbf{0} & \cdots  & {{M}_{33}} & M_{34}^{*}  \\
        \mathbf{0} & \mathbf{0} & \mathbf{0} & \cdots  & M_{35}^{*} & M_{33}^{*}  \\
        \end{pmatrix}.
      \end{equation}

      \begin{equation}
        {M'_{3}}=\begin{pmatrix}
            {M'_{30}} & {M'_{31}} & \mathbf{0} & \cdots  & \mathbf{0} & \mathbf{0}  \\
            {M'_{32}} & {M'_{33}} & {M'_{34}} & \cdots  & \mathbf{0} & \mathbf{0}  \\
            \mathbf{0} & {M'_{35}} & {M'_{33}} & \cdots  & \mathbf{0} & \mathbf{0}  \\
            \vdots  & \vdots  & \vdots  & \ddots  & \vdots  & \vdots   \\
            \mathbf{0} & \mathbf{0} & \mathbf{0} & \cdots  & {M'_{33}} & M'^{*}_{34}  \\
            \mathbf{0} & \mathbf{0} & \mathbf{0} & \cdots  & M'^{*}_{35} & M'^{*}_{33}  \\
        \end{pmatrix}.
      \end{equation}

      \begin{equation}
          {{M}_{4}}=\begin{pmatrix}
           {{M}_{40}} & \mathbf{0} & \mathbf{0} & \cdots  & \mathbf{0} & \mathbf{0}  \\
           {{M}_{41}} & {{M}_{42}} & \mathbf{0} & \cdots  & \mathbf{0} & \mathbf{0}  \\
           0 & {{M}_{43}} & {{M}_{42}} & \cdots  & \mathbf{0} & \mathbf{0}  \\
           \vdots  & \vdots  & \vdots  & \ddots  & \vdots  & \vdots   \\
           \mathbf{0} & \mathbf{0} & \mathbf{0} & \cdots  & {{M}_{42}} & \mathbf{0}  \\
           \mathbf{0} & \mathbf{0} & \mathbf{0} & \cdots  & M_{43}^{*} & M_{42}^{*}  \\
          \end{pmatrix}.
      \end{equation}

      \begin{equation}
        {{M}_{5}}=\begin{pmatrix}
            {{\mathbf{0}}_{\tau_1\times\tau_1}} & {{M}_{50}} & \mathbf{0} & \cdots  & \mathbf{0} & \mathbf{0}  \\
            \mathbf{0} & {{M}_{51}} & {{M}_{52}} & \cdots  & \mathbf{0} & \mathbf{0}  \\
            \mathbf{0} & \mathbf{0} & {{M}_{51}} & \cdots  & \mathbf{0} & \mathbf{0}  \\
            \vdots  & \vdots  & \vdots  & \ddots  & \vdots  & \vdots   \\
            \mathbf{0} & \mathbf{0} & \mathbf{0} & \cdots  & {{M}_{51}} & M_{52}^{*}  \\
            \mathbf{0} & \mathbf{0} & \mathbf{0} & \cdots  & \mathbf{0} & M_{51}^{*}  \\
        \end{pmatrix}.
      \end{equation}

\begin{remark}
  In this paper, compared with the model in \cite{Alfa-1995}, two components are added to the state of Markov chain, namely, the length of queue 2 and the phase of the D-PH distribution followed by the service time of each task in queue 2; moreover the state space is finite. Compared with the model in \cite{Baumann-2017}, the phases of the D-PH distribution followed by the vacation duration of server 1 are considered. These differences distinguish the delay analysis.
\end{remark}

\subsection{The stationary distribution of the Markov chain $\Xi$}

The matrix-geometric method proposed by Neuts \cite{Neuts-73} has been widely used for the exact analysis of queueing systems that have an embedded Markov chain with block tridiagonal transition matrix. For instance, Liu \MakeLowercase{\textit{et al.}} \cite{LiuJ-2012} proposed an opportunistic scheduling scheme, MARCH, to improve the QoS of secondary users in cognitive radio systems. Then, they formulated a tandem queueing system with an infinite dimensional block tridiagonal transition matrix and analyzed the performance of MARCH based on the matrix-geometric method. Kumar \MakeLowercase{\textit{et al.}} \cite{KumarW-2014} analyzed the performance of an energy-efficient motorway vehicular communication system by formulating two different queueing models, which have an infinite and finite dimensional block tridiagonal transition matrix respectively, based on the matrix-geometric method. According to the structural feature of the one-step transition probability matrix $P$, the matrix-geometric method can be used to analyze the stationary distribution of the Markov chain $\Xi$.

Let $\mathbf{x}$ denote the stationary distribution of the Markov chain $\Xi$:
\begin{equation}
  \mathbf{x}=\begin{pmatrix}
   {{\mathbf{x}}_{0}} & {{\mathbf{x}}_{1}} & {{\mathbf{x}}_{2}} & \cdots  & {{\mathbf{x}}_{{{N}_{1}}}}  \\
\end{pmatrix},
\end{equation}
where
	\begin{align}
      &  {{\mathbf{x}}_{{{i}_{1}}}}=\begin{pmatrix}
           {{\mathbf{x}}_{{{i}_{1}},0}} & {{\mathbf{x}}_{{{i}_{1}},1}} & \cdots  & {{\mathbf{x}}_{{{i}_{1}},{{N}_{2}}}}  \\
         \end{pmatrix},\\
      &  {{\mathbf{x}}_{{{i}_{1}},{{i}_{2}}}}=\left( {{\left( {{\mathbf{x}}_{{{i}_{1}},{{i}_{2}}}} \right)}_{\bm{\varphi }}}:\bm{\varphi }\in {{\Phi }_{{{i}_{1}},{{i}_{2}}}} \right), \nonumber\\
      &  \phantom{{{\mathbf{x}}_{{{i}_{1}},{{i}_{2}}}}={{\left( {{\mathbf{x}}_{{{i}_{1}},{{i}_{2}}}} \right)}_{\bm{\varphi }}}:} 0\le {{i}_{1}}\le {{N}_{1}},\; 0\le {{i}_{2}}\le {{N}_{2}}.
    \end{align}
The element ${{\left( {{\mathbf{x}}_{{{i}_{1}},{{i}_{2}}}} \right)}_{\bm{\varphi }}}$ represents the probability that the state of the tandem queueing system is $\left( {{i}_{1}},{{i}_{2}},\bm{\varphi } \right)$ at times ${{t}_{0}}$, ${{t}_{1}}$, ${{t}_{2}}$, $\cdots $. The stationary distribution ${\mathbf{x}}$ satisfies the following system of equations:
\begin{equation} \label{eq-StaDis}
  \begin{cases}
  \mathbf{x}P=\mathbf{x} \\
  \mathbf{xe}=1 \\
 \end{cases}.
\end{equation}
In \eqref{eq-StaDis}, the first equation represents that given the Markov chain $\Xi$ with the stationary distribution as the initial state distribution, the distribution at arbitrary time will be the same as the initial distribution; the second equation represents that the state transitions are always performed in the state space ${\Psi }$. From the above system of equations, the stationary distribution ${\mathbf{x}}$ can be obtained through the method in Appendix \ref{Appendix-C}.

Let $P_{2,full}$ denote the probability that the residual computation buffer is not enough for one task, then $P_{2,full}$ can be calculated as
\begin{equation}
  {P}_{2,full}={{\mathbf{x}}_{0}}{{\mathbf{e}}_{1}}+\sum\limits_{i=1}^{{{N}_{1}}}{{{\mathbf{x}}_{i}}{{\mathbf{e}}_{2}}}, \label{P_2full}
\end{equation}
where
\begin{align}
  & {{\mathbf{e}}_{1}}={{\begin{pmatrix}
    {{\mathbf{0}}_{1\times \left( m+\tau_4\left( {{N}_{2}}-1 \right) \right)}} & {{\mathbf{1}}_{1\times \tau_4}}  \\
    \end{pmatrix}}^{T}}, \\
  & {{\mathbf{e}}_{2}}={{\begin{pmatrix}
    {{\mathbf{0}}_{1\times \left[ \tau_1+\left( \tau_2+\tau_5\right)\left( {{N}_{2}}-1 \right) \right]}} & {{\mathbf{1}}_{1\times \tau_2}}  \\
    \end{pmatrix}}^{T}}.
\end{align}
In \eqref{P_2full}, the first item ${{\mathbf{x}}_{0}}{{\mathbf{e}}_{1}}$ denotes the probability that there are $0$ and $N_2$ tasks in queue 1 and queue 2, respectively, i.e., ${{\mathbf{x}}_{0}}{{\mathbf{e}}_{1}}=\mathbf{x}_{0,{N_{2}}}\mathbf{e}$. In the second item of \eqref{P_2full}, the basic element ${{\mathbf{x}}_{i}}{{\mathbf{e}}_{2}}$ denotes the probability that there are $i$ and $N_2$ tasks in queue 1 and queue 2, respectively, i.e., ${{\mathbf{x}}_{i}}{{\mathbf{e}}_{2}}=\mathbf{x}_{i,{N_{2}}}\mathbf{e}$, $i=1,2,\cdots,N_1$. The saturation of the computation buffer affects the transmission waiting time and the computation waiting time. Hence, the variation of $P_{2,full}$ will be used as a reference to analyze the reason for the variations in delay characteristics with the task transmission rate in Section \ref{sec-52}.

\subsection{Estimation methods for the random delay characteristics}

The cumulative probability distribution (CPD) of the delay is defined by ${\bar{W}_{n}}=P\{T\leq n\}$, where $T$ denotes the delay considered in this paper and $n$ $(n\ge 1)$ denotes the number of time slots. The probability distribution of the delay is defined by ${{PW}_{n}}=P\{T=n\}$. The delay bound violation probability is defined by ${{W}_{n}}=P\{T > n\}$.

Given the initial state that follows the event that a task enters queue 1, let
\begin{equation}
  	{{\mathbf{\tilde{x}}}^{(n)}}=\begin{pmatrix}
    \mathbf{\tilde{x}}_{0}^{(n)} & \mathbf{\tilde{x}}_{1}^{(n)} & \mathbf{\tilde{x}}_{2}^{(n)} & \cdots  & \mathbf{\tilde{x}}_{{{N}_{1}}}^{(n)} \\
    \end{pmatrix}, \label{eq-xtilde}
\end{equation}
where
\begin{align}
  & \mathbf{\tilde{x}}_{{{i}_{1}}}^{(n)}=\begin{pmatrix}
    \mathbf{\tilde{x}}_{{{i}_{1}}0}^{(n)} & \mathbf{\tilde{x}}_{{{i}_{1}}1}^{(n)} & \cdots  & \mathbf{\tilde{x}}_{{{i}_{1}}{{N}_{2}}}^{(n)}  \\
    \end{pmatrix}, \label{eq-xtilde1}\\
  & \mathbf{\tilde{x}}_{{{i}_{1}}{{i}_{2}}}^{(n)}=\left( {{\left( \mathbf{\tilde{x}}_{{{i}_{1}}{{i}_{2}}}^{(n)} \right)}_{\bm{\varphi }}}:\bm{\varphi }\in {{\Phi }_{{{i}_{1}},{{i}_{2}}}} \right),\nonumber\\
  & \phantom{\mathbf{\tilde{x}}_{{{i}_{1}}{{i}_{2}}}^{(n)}={{\left( \mathbf{\tilde{x}}_{{{i}_{1}}{{i}_{2}}}^{(n)} \right)}_{\mathbf{\varphi }}}:} 0\le {{i}_{1}}\le {{N}_{1}},\;0\le {{i}_{2}}\le {{N}_{2}},  \label{eq-xtilde2}
\end{align}
and ${{\left( \mathbf{\tilde{x}}_{{{i}_{1}}{{i}_{2}}}^{(n)} \right)}_{\bm{\varphi }}}$ represents the probability that the state of the tandem queueing system is $\left( {{i}_{1}},{{i}_{2}},\bm{\varphi } \right)$ after $n$ time slots. During these $n$ time slots, assume that the tasks entering queue 1 are not counted in the queue length.

\begin{proposition} \label{prop-PD}
    The cumulative probability distribution of the delay can be calculated as
    \begin{equation}
      \bar{W}_{n}=\mathbf{\tilde{x}}_{0\text{0}}^{(n)}\mathbf{e}, \quad n\ge 1; \label{eq-CPD}
    \end{equation}
    and the probability distribution of the delay can be calculated as
    \begin{equation}
      {{PW}_{n}}={\bar{W}_{n}}-{\bar{W}_{n-1}}, \quad n\ge 1,  \label{eq-PD}
    \end{equation}
    where ${\bar{W}_{0}=0}$. The delay bound violation probability is calculated as
    \begin{equation}
      {{W}_{n}}=1-\bar{W}_{n},\quad n\ge 1. \label{equ-delayBVP}
    \end{equation}
\end{proposition}
\begin{IEEEproof}
  According to \eqref{eq-xtilde}, \eqref{eq-xtilde1} and \eqref{eq-xtilde2}, $\mathbf{\tilde{x}}_{0\text{0}}^{(n)}\mathbf{e}$ represents the probability that after $n$ time slots from the instant when a task enters queue 1, the task and the ones queued ahead of it will all have departed from queue 2. Therefore, equation \eqref{eq-CPD} is obtained. Since $PW_{n}$ denotes the probability that the delay of the task is $n$ time slots, i.e., $PW_{n}=P\{T=n\}$, and $\bar{W}_{n}=P\{T\leq n\}$, $PW_{n}$ can be calculated as equation \eqref{eq-PD}. Since ${{W}_{n}}$ denotes the probability that the delay of the task exceeds $n$ time slots, i.e., ${{W}_{n}}=P\{T>n\}$, it can be calculated as equation \eqref{equ-delayBVP}.
\end{IEEEproof}

For $\mathbf{\tilde{x}}_{0\text{0}}^{(n)}$ in \eqref{eq-CPD}, it can be calculated by the following iterative method.
    \begin{align}
      & \mathbf{\tilde{x}}_{0}^{(n)}=\mathbf{\tilde{x}}_{0}^{(n\text{-1})}{{\tilde{M}}_{0}}+\mathbf{\tilde{x}}_{1}^{(n\text{-1})}{{\tilde{M}}_{2}}, \\
      & \mathbf{\tilde{x}}_{1}^{(n)}=\mathbf{\tilde{x}}_{0}^{(n\text{-1})}{{\tilde{M}}_{1}}+\mathbf{\tilde{x}}_{1}^{(n\text{-1})}{{\tilde{M}}_{3}}+\mathbf{\tilde{x}}_{2}^{(n\text{-1})}{{\tilde{M}}_{5}},\\
      & \mathbf{\tilde{x}}_{{{i}_{2}}}^{(n)}=\mathbf{\tilde{x}}_{{{i}_{2}}-1}^{(n\text{-1})}{{\tilde{M}}_{4}}+\mathbf{\tilde{x}}_{{{i}_{2}}}^{(n\text{-1})}{{\tilde{M}}_{3}}+\mathbf{\tilde{x}}_{{{i}_{2}}+1}^{(n\text{-1})}{{\tilde{M}}_{5}}, \nonumber\\
      & \phantom{\mathbf{\tilde{x}}_{{{i}_{2}}}^{(n)}=\mathbf{\tilde{x}}_{{{i}_{2}}-1}^{(n\text{-1})}{{\tilde{M}}_{4}}+\mathbf{\tilde{x}}_{{{i}_{2}}}^{(n\text{-1})}} 2\le {{i}_{2}}\le {{N}_{1}}-1, \\
      & \mathbf{\tilde{x}}_{{{N}_{1}}}^{(n)}=\mathbf{\tilde{x}}_{{{N}_{1}}-1}^{(n\text{-1})}{{\tilde{M}}_{4}}+\mathbf{\tilde{x}}_{{{N}_{1}}}^{(n\text{-1})}\tilde{M}_{3}',
    \end{align}
where ${{\tilde{M}}_{i}}$ $(i=0,1,\cdots,5)$ and $\tilde{M}_{3}'$ are obtained by replacing $D_0$ and $D_1$ in the expressions of the elements in ${{M}_{i}}$ $(i=0,1,\cdots,5)$ and $M_{3}'$ with $\mathbf{I}$, respectively, except that the matrix blocks $M_{00}$ and $M_{10}$ are set to $\mathbf{I}$ and $\mathbf{0}$, respectively. The initial input ${{\mathbf{\tilde{x}}}^{(0)}}$ of the above iterative method is the stationary distribution of the tandem queueing system just after a task enters queue 1. The initial input ${{\mathbf{\tilde{x}}}^{(0)}}$ can be calculated as follows. For simplifying the notations, let $\mathbf{z}={{\mathbf{\tilde{x}}}^{(0)}}$.

\begin{equation}
  \mathbf{z}=\begin{pmatrix}
   {{\mathbf{z}}_{0}} & {{\mathbf{z}}_{1}} & {{\mathbf{z}}_{2}} & \cdots  & {{\mathbf{z}}_{{{N}_{1}}}}  \\
  \end{pmatrix},
\end{equation}
where
\begin{align}
      & {{\mathbf{z}}_{0}}=\mathbf{0};\\
      &  {{\mathbf{z}}_{{{i}_{1}}}}=\begin{pmatrix}
           {{\mathbf{z}}_{{{i}_{1}}0}} & {{\mathbf{z}}_{{{i}_{1}}1}} & \cdots  & {{\mathbf{z}}_{{{i}_{1}}{{N}_{2}}}}  \\
         \end{pmatrix},\\
      &  {{\mathbf{z}}_{{{i}_{1}}{{i}_{2}}}}=\left( {{\left( {{\mathbf{z}}_{{{i}_{1}}{{i}_{2}}}} \right)}_{\bm{\varphi }}}:\bm{\varphi }\in {{\Phi }_{{{i}_{1}},{{i}_{2}}}} \right),\nonumber\\
      &  \phantom{{{\mathbf{z}}_{{{i}_{1}}{{i}_{2}}}}={{\left( {{\mathbf{z}}_{{{i}_{1}}{{i}_{2}}}} \right)}_{\bm{\varphi }}}:} 0\le {{i}_{1}}\le {{N}_{1}},\; 0\le {{i}_{2}}\le {{N}_{2}},
    \end{align}
and ${{\left( {{\mathbf{z}}_{{{i}_{1}}{{i}_{2}}}} \right)}_{\bm{\varphi }}}$ represents the probability that the state of the tandem queueing system is $\left( {{i}_{1}},{{i}_{2}},\bm{\varphi } \right)$ just after a task enters queue 1. The initial input ${\mathbf{z}}$ can be calculated as
\begin{equation}
   \mathbf{z}=\frac{\begin{pmatrix}
    {{{\mathbf{\hat{z}}}}_{0}} & {{{\mathbf{\hat{z}}}}_{1}} & {{{\mathbf{\hat{z}}}}_{2}} & \cdots  & {{{\mathbf{\hat{z}}}}_{{{N}_{1}}}}  \\
    \end{pmatrix}}{\begin{pmatrix}
       {{{\mathbf{\hat{z}}}}_{0}} & {{{\mathbf{\hat{z}}}}_{1}} & {{{\mathbf{\hat{z}}}}_{2}} & \cdots  & {{{\mathbf{\hat{z}}}}_{{{N}_{1}}}}  \\
    \end{pmatrix}\mathbf{e}},
\end{equation}
where
\begin{align}
  & {{\mathbf{\hat{z}}}_{0}}={{\mathbf{x}}_{0}}{{\hat{M}}_{0}}+{{\mathbf{x}}_{1}}{{\hat{M}}_{2}},\\
  & {{\mathbf{\hat{z}}}_{1}}={{\mathbf{x}}_{0}}{{\hat{M}}_{1}}+{{\mathbf{x}}_{1}}{{\hat{M}}_{3}}+{{\mathbf{x}}_{2}}{{\hat{M}}_{5}},\\
  & {{\mathbf{\hat{z}}}_{i}}={{\mathbf{x}}_{i-1}}{{\hat{M}}_{4}}+{{\mathbf{x}}_{i}}{{\hat{M}}_{3}}+{{\mathbf{x}}_{i+1}}{{\hat{M}}_{5}}, \quad 2\le i\le {{N}_{1}}-1,\\
  & {{\mathbf{\hat{z}}}_{{{N}_{1}}}}={{\mathbf{x}}_{{{N}_{1}}-1}}{{\hat{M}}_{4}}+{{\mathbf{x}}_{{{N}_{1}}}}\hat{M}_{3}'.
\end{align}
In the above equations, ${{\hat{M}}_{i}}$ $(i=0,1,\cdots,5)$ and $\hat{M}_{3}'$ are equal to ${{M}_{i}}$ $(i=0,1,\cdots,5)$ and ${M}_{3}'$, respectively, except that the matrix elements that correspond to the one-step transitions without any task entering queue 1 are set to 0.

Let ${P_{off}}$ denote the offloading ratio, which is the probability that the task can enter queue 1 after it arrives at queue 1.
  \begin{equation}
      {{P}_{off}}={{\lambda }^{-1}}\left( \begin{matrix}
       {{{\mathbf{\hat{z}}}}_{0}} & {{{\mathbf{\hat{z}}}}_{1}} & {{{\mathbf{\hat{z}}}}_{2}} & \cdots  & {{{\mathbf{\hat{z}}}}_{{{N}_{1}}}}  \\
      \end{matrix} \right)\mathbf{e}.
  \end{equation}
The offloading ratio can measure the number of tasks entering the MEC system per time slot. Namely, the increase of ${P_{off}}$ implies that more tasks can enter the MEC system per time slot. The number of tasks in the system affects the transmission waiting time and the computation waiting time. Thus, the variation of ${P_{off}}$ will be used as a reference to analyze the reason for the variations in delay characteristics with the task transmission rate in Section \ref{sec-52}.

\begin{proposition} \label{prop-ave}
  The average delay $D_{ave}$ can be calculated by either of the following two methods:
  \begin{gather}
      {{D}_{ave}}=\sum\limits_{n=1}^{\infty }{nPW_{n}}; \label{equ-ave1}\\
      {{D}_{ave}}=\frac{\sum\limits_{{{i}_{1}}=0}^{{{N}_{1}}}{\sum\limits_{{{i}_{2}}=0}^{{{N}_{2}}}{\left( {{i}_{1}}+{{i}_{2}} \right){{\mathbf{x}}_{{{i}_{1}}{{i}_{2}}}}\mathbf{e}}}}{\lambda{P_{off}}}. \label{equ-ave2}
  \end{gather}
\end{proposition}
\begin{IEEEproof}
  Equation \eqref{equ-ave1} is obtained based on the definition of the average delay. Equation \eqref{equ-ave2} is obtained according to Little's Law.
\end{IEEEproof}

\begin{proposition} \label{prop-sd}
  The delay standard deviation $D_{sd}$ can be calculated as
  \begin{equation}
      {{D}_{sd}}=\sqrt{\sum\limits_{n=1}^{\infty }{PW_{n}{{\left( n-{{D}_{ave}} \right)}^{2}}}}. \label{equ-sd}
  \end{equation}
\end{proposition}
\begin{IEEEproof}
    Equation \eqref{equ-sd} is obtained based on the definition of the delay standard deviation.
\end{IEEEproof}

\begin{remark}
    In the above three propositions, the cumulative probability distribution $\bar{W}_{n}$ of the delay determines all of the concerned random delay characteristics, and it relies on the evolution of the number of tasks in the MEC system. From the perspective of task number evolution, the MEC system can be represented by a Markov chain. The one-step transition probability matrix of the Markov chain reflects the task number evolution mechanism, which is formed by the interaction of the task generation process, the transmission process and the computation process. From the construction of the transition probability matrix and the derivation of $\bar{W}_{n}$, the influence of any process on the delay characteristics is nonlinear and complex.
\end{remark}

\subsection{Discussion on the multi-user case}

In the multi-user case, the delay characteristics of different users are coupled with each other. If the computations of different users are executed independently in their own VMs of the MEC server, the mutual influences mainly come from the wireless resource sharing among different users. Thus, the task transmission time distribution should be modified accordingly, and characterize the delay characteristics of multiple users. Toward this end, we choose two typical multi-access strategies, i.e., Orthogonal Frequency Division Multiple Access (OFDMA) and Non-orthogonal Multiple Access (NOMA), to briefly demonstrate the extension of this work to the multi-user case.

Suppose that $N$ users share the uplink bandwidth $B$ and the task size of UE $i$ $(i=1,2,\cdots,N)$ is $L_i$. For the task transmission time distribution\footnote{The task transmission time of UE $i$ obeys the D-PH distribution $\left(\beta _{1}^{i},S_{1}^{i}\right)$.} of UE $i$, its parameters $\left(\beta _{1}^{i},S_{1}^{i}\right)$ are obtained by fitting the D-PH distribution to the sampling values of task transmission times. The task transmission time $R_{i}$ can be sampled according to the formula $R_{i}={{L}_{i}}/{r_{i}}$, where ${r_{i}}$ denotes the data transmission rate. For instance, if OFDMA is applied, $r_{i}$ can be calculated as
\begin{equation}
  r_{i}=\left( B-\sum\limits_{n=1,n\ne i}^{N}{{{B}_{n}}} \right){{\log }_{2}}\left( 1+\frac{{p_i}{{\left| {{h}_{i}} \right|}^{2}}}{{{\sigma }_i}} \right); \label{equ-OFDMA1}
\end{equation}
if NOMA is applied, $r_{i}$ can be calculated as
\begin{equation}
  r_{i}=B{{\log }_{2}}\left( 1+\frac{{{p}_{i}}{{\left| {{h}_{i}} \right|}^{2}}}{{{\sigma }_i}+\sum\limits_{n=1,n\neq i}^{N}{{{p}_{n}}{{\left| {{h}_{n}} \right|}^{2}}}} \right). \label{equ-NOMA1}
\end{equation}
In equations \eqref{equ-OFDMA1} and \eqref{equ-NOMA1}, $p_i$, ${{\left| {{h}_{i}} \right|}^{2}}$ and $\sigma_i$ represent the transmission power, the channel power gain and the background white Gaussian noise power of UE $i$, respectively. Then, the holistic random delay characteristics of UE $i$ can be estimated according to the methods derived in the previous subsection. The connection between the parameters $\left(\beta _{1}^{i},S_{1}^{i}\right)$ of UE $i$ and the existence of other users is reflected implicitly in fitting the D-PH distribution to the task transmission times. It is shown in \eqref{equ-OFDMA1} that the effect of other users on the delay characteristics of UE $i$ is caused by the wireless bandwidth allocation policy. It is shown in \eqref{equ-NOMA1} that the effect of other users on the delay characteristics of UE $i$ is caused by the interference from transmission powers and channel power gains.

If different users share the computation resources in the same VM, the influences among the delay characteristics of different users will be more complex. The influence mechanism is worth studying in future works.

\section{Experiments} \label{sec-5}

In this section, the effectiveness of the estimation methods for the random delay characteristics is verified first. Then, the experimental analysis of the random delay characteristics is carried out.

\subsection{Effectiveness of the estimation methods for the random delay characteristics} \label{sec-51}

The effectiveness of the estimation methods for the random delay characteristics will be verified through comparison with the simulation method. There are four random delay characteristics. According to Proposition \ref{prop-PD}, Proposition \ref{prop-ave} and Proposition \ref{prop-sd}, it is enough to prove the effectiveness of the estimation methods by just verifying the delay bound violation probability and the average delay. The simulation is implemented using the sequential procedure. During the implementation of the sequential procedure, four simulators run simultaneously. They are used for generating tasks according to the given D-MAP and for sampling the transmission time of each task, the computation time of each task and the vacation duration according to the given respective D-PH distributions. The delay of each task is recorded. From these records, the delay bound violation probability and the average delay can be calculated. The final results are outputted based on the termination condition of the sequential procedure, i.e., the confidence level is set to $95\%$ and the relative accuracy is set to $0.05$.

For the delay bound violation probability, we will compare the values obtained by the simulation method and equation \eqref{equ-delayBVP} (called the method based on the CPD) in two cases with the following parameters.

\begin{enumerate}
  \item Case 1: the average transmission rate is less than the average computation rate, i.e., $\mu_1<\mu_2$; the average duration of one transmission pause is shorter and it is set to $2\Delta t$.

  The task arrival process is defined by matrices $D_0$ and $D_1$ as
    \begin{equation*}
      {{D}_{0}}=\begin{pmatrix}
       0.2359 & 0.1938  \\
       0.2792 & 0.2805  \\
      \end{pmatrix},\;
      {{D}_{1}}=\begin{pmatrix}
       0.1236 & 0.4467  \\
       0.2644 & 0.1759  \\
      \end{pmatrix},
    \end{equation*}
    where $\lambda=0.5$ task$/\Delta t$. The transmission time of each task obeys a D-PH distribution with the representation $({{\beta }_{1}},{{S}_{1}})$, where ${{\beta }_{1}}=1$, ${{S}_{1}}=0.6429$ and ${{\mu }_{1}}=0.3571$ task$/\Delta t$. The vacation duration of server 1 obeys a D-PH distribution with the representation $(v,V)$, where $v=\begin{pmatrix}
   0.6545 & 0.3455  \\
  \end{pmatrix} $, $V=\begin{pmatrix}
   0.3035 & 0.0617  \\
   0.6738 & 0.1916  \\
  \end{pmatrix}$ and ${{\eta }_{2}}=2\Delta t$. The computation time of each task obeys a D-PH distribution with the representation $({{\beta }_{2}},{{S}_{2}})$, where ${{\beta }_{2}}=1$, ${{S}_{2}}=0.5455$ and ${{\mu }_{2}}=0.4545$ task$/\Delta t$. The transmission buffer and computation buffer can accommodate 10 and 15 tasks, respectively, i.e., $N_1=10$ and $N_2=15$. Let $\Delta t=1$ ms.

  \item Case 2: the average transmission rate is greater than the average computation rate, i.e., $\mu_1>\mu_2$; the average duration of one transmission pause is longer and it is set to $4\Delta t$.

  The parameters are the same as case 1, except for the following. The transmission time of each task obeys a D-PH distribution with the representation $({{\beta }_{1}},{{S}_{1}})$, where ${{\beta }_{1}}=1$, ${{S}_{1}}=0.1667$ and ${{\mu }_{1}}=0.8333$ task$/\Delta t$. The vacation duration of server 1 obeys a D-PH distribution with the representation $(v,V)$, where $v=\begin{pmatrix}
   0.6969 & 0.3031  \\
  \end{pmatrix} $, $V=\begin{pmatrix}
   0.6378 & 0.1007  \\
   0.4613 & 0.3278  \\
  \end{pmatrix}$ and ${{\eta }_{2}}=4\Delta t$.
\end{enumerate}

\begin{figure}[!t]
\centering
\includegraphics[width=3.4in]{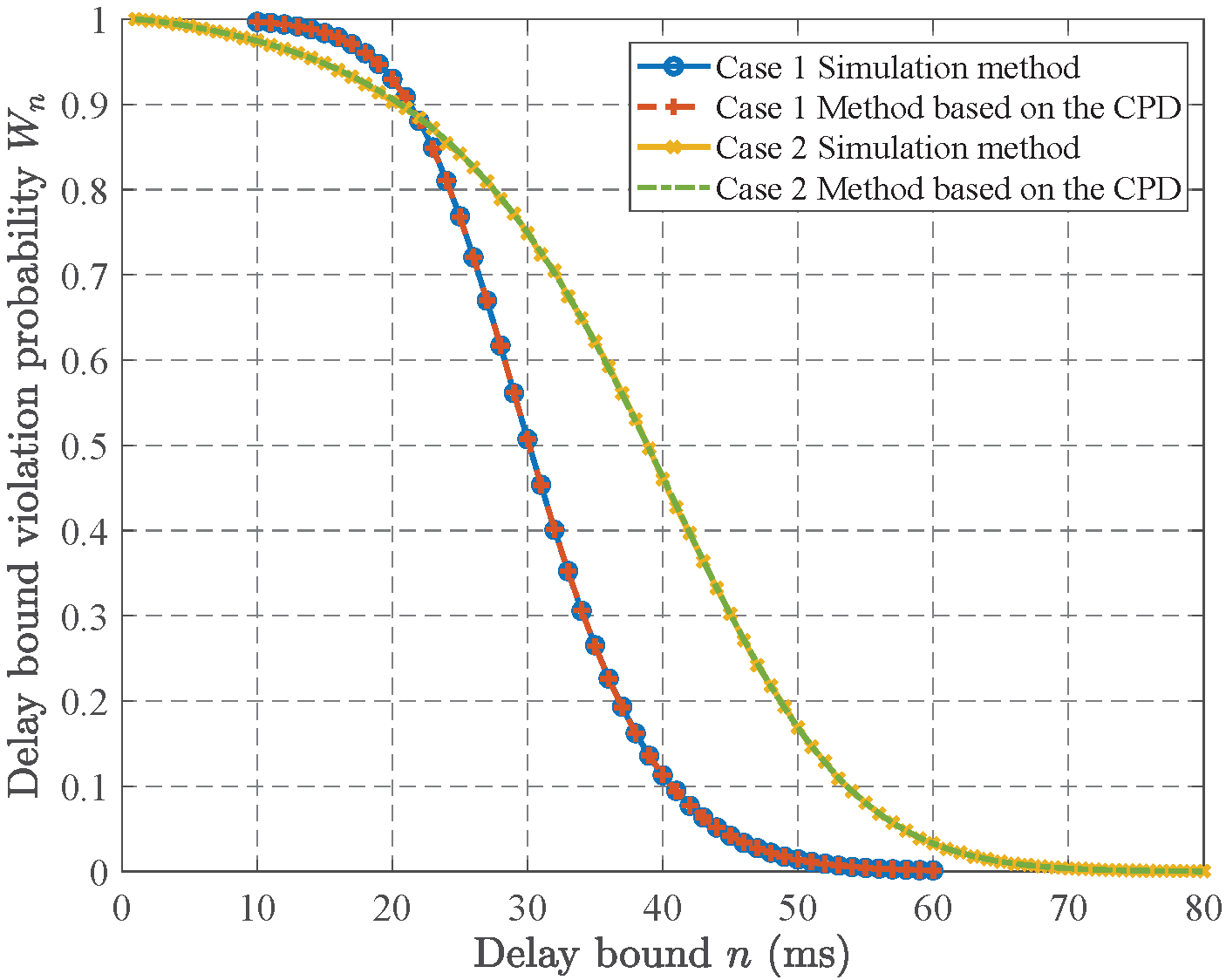}
\caption{The delay bound violation probability $W_n$ vs. the delay bound $n$.}
\label{fig-04}
\end{figure}

\begin{table*}
    \renewcommand{\arraystretch}{1.3}
    \caption{Delay bound violation probabilities obtained through different methods for case 1}
    \label{Tab-3}
    \centering
    \begin{tabular}{c|ccc}
    \hline
    \multirow{4}{*}{Delay bound (ms)} & \multicolumn{3}{c}{Delay bound violation probability}                                                                                                                     \\ \cline{2-4}
                                      & \multirow{2}{*}{Method based on the CPD} & \multicolumn{2}{c}{\begin{tabular}[c]{@{}c@{}}Simulation method\\ (Confidence level: 95\%)\end{tabular}} \\ \cline{3-4}
                                      &                                          & Value                                                   & Confidence interval                                                  \\ \hline
    10                                & 0.9970                                   & 0.9970                                                  & {[}0.9969, 0.9971{]}                                                  \\
    20                                & 0.9296                                   & 0.9300                                                  & {[}0.9293, 0.9307{]}                                                  \\
    30                                & 0.5072                                   & 0.5073                                                  & {[}0.5055, 0.5091{]}                                                  \\
    40                                & 0.1132                                   & 0.1131                                                  & {[}0.1121, 0.1141{]}                                                  \\
    50                                & 0.0139                                   & 0.0140                                                  & {[}0.0137, 0.0143{]}                                                  \\
    60                                & 0.0012                                   & 0.0012                                                  & {[}0.0011, 0.0013{]}                                                  \\ \hline
    \end{tabular}
\end{table*}

\begin{table*}
    \renewcommand{\arraystretch}{1.3}
    \caption{Delay bound violation probabilities obtained through different methods for case 2}
    \label{Tab-4}
    \centering
    \begin{tabular}{c|ccc}
    \hline
    \multirow{4}{*}{Delay bound (ms)} & \multicolumn{3}{c}{Delay bound violation probability}                                                                                                                     \\ \cline{2-4}
                                      & \multirow{2}{*}{Method based on the CPD} & \multicolumn{2}{c}{\begin{tabular}[c]{@{}c@{}}Simulation method\\ (Confidence level: 95\%)\end{tabular}} \\ \cline{3-4}
                                      &                                          & Value                                                   & Confidence interval                                                  \\ \hline
    10                                & 0.9745                                   & 0.9739                                                  & {[}0.9730, 0.9748{]}                                                  \\
    20                                & 0.9064                                   & 0.9053                                                  & {[}0.9035, 0.9071{]}                                                  \\
    30                                & 0.7496                                   & 0.7487                                                  & {[}0.7460, 0.7514{]}                                                  \\
    40                                & 0.4631                                   & 0.4623                                                  & {[}0.4599, 0.4647{]}                                                  \\
    50                                & 0.1687                                   & 0.1696                                                 & {[}0.1681, 0.1711{]}                                                  \\
    60                                & 0.0327                                   & 0.0328                                                  & {[}0.0323, 0.0333{]}                                                  \\
    70                                & 0.0035                                   & 0.0036                                                  & {[}0.0035, 0.0037{]}                                                  \\
    80                                & 0.0002                                   & 0.0002                                                  & {[}0.00017,0.00023{]}                                                  \\ \hline
    \end{tabular}
\end{table*}

 For each case, the variation in the delay bound violation probability $W_n$ with the delay bound $n$ is illustrated in Fig. \ref{fig-04}. It is shown that the delay bound violation probabilities obtained by the simulation method and the method based on the CPD are closely consistent. From Table \ref{Tab-3} and Table \ref{Tab-4}, the delay bound violation probabilities obtained through the estimation method all lie within the confidence intervals with the confidence level 95\%. In addition, it is known that when the average transmission rate is larger, more tasks can enter the MEC system, this will degrade the delay performance; moreover, the longer vacation time will also degrade the delay performance. These are consistent with the experimental results shown in Fig. \ref{fig-04}. Namely, for case 1, the delay mostly falls in the interval between 10 ms and 60 ms; for case 2, the delay mostly falls in the interval between 1 ms and 80 ms; moreover, the former case has a shorter tail.

 \begin{table*}[!t]
  \renewcommand{\arraystretch}{1.3}
  \caption{The task transmission rate $\mu_1$ and its corresponding D-PH distribution $\left({{\beta }_{1}},S_1\right)$}
  \label{Tab-1}
  \centering
  \begin{tabular}{c|cccccccccc}
  \hline
  $\mu_1$ & 0.1429 & 0.1786 & 0.2381 & 0.3571 & 0.4 & 0.4545 & 0.5263 & 0.625 & 0.7692 & 0.8333 \\
  \hline
  ${{\beta }_{1}}$ & 1 & 1 & 1 & 1 & 1 & 1 & 1 & 1 & 1 & 1 \\
  $S_1$ & 0.8571 & 0.8214 & 0.7619 & 0.6429 & 0.6 & 0.5455 & 0.4737 & 0.375 & 0.2308 & 0.1667 \\
  \hline
\end{tabular}
\end{table*}

\begin{figure}[!t]
\centering
\includegraphics[width=3.4in]{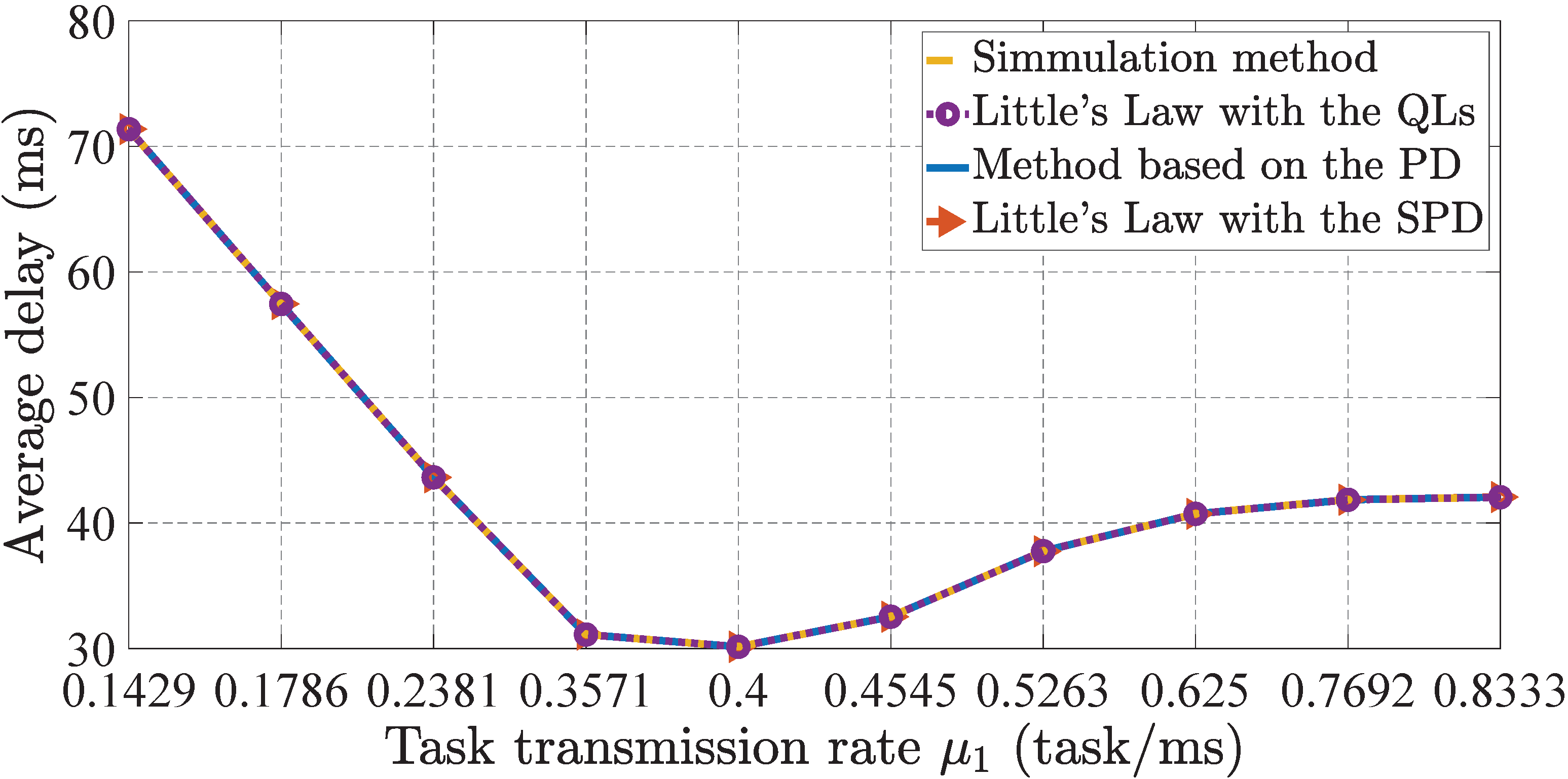}
\caption{The average delay $D_{ave}$ vs. the task transmission rate $\mu_1$, where the values of the average delay are obtained by four methods.}
\label{fig-05}
\end{figure}

Consider case 1 again. We observe the variation in the average delay with the task transmission rate ${{\mu }_{1}}$ in Table \ref{Tab-1}. The average delay is obtained by the simulation method, the method in \cite{MaoY-2017,KimY-2018,Meng-2019}, equation \eqref{equ-ave1} and equation \eqref{equ-ave2} in this paper, respectively. The method in \cite{MaoY-2017,KimY-2018,Meng-2019} is based on Little's Law with queue lengths (QLs) over an infinite time horizon; equation \eqref{equ-ave1} is based on the PD of the delay, and equation \eqref{equ-ave2} is based on Little's Law with the stationary probability distribution (SPD) of queue lengths. It is shown in Fig. \ref{fig-05} that the average delays obtained by the above four methods are closely consistent.

It is shown from the above experiments that the estimation methods in this paper are effective.

\begin{remark}
    It is shown in Table \ref{Tab-3} and Table \ref{Tab-4} that the results obtained by the estimation method and the simulation method are very similar. The reasons are analyzed as follows. First, the theoretical derivation of the estimation method does not involve any approximate substitution. Second, in the simulation, the termination condition of the sequential process makes the simulation carry out a large number of independent repeated experiments. Third, there are similarities between the estimation method and the simulation method as follows. In the estimation method, the state transition process of the system is dominated by the one-step transition probability matrix of the Markov chain. This transition probability matrix is derived theoretically, based on the task generation process, the transmission time distribution of each task, the vacation duration distribution of the transmitter and the computation time distribution of each task. In the simulation method, the state transition process of the system is dominated by four simulators. These simulators are responsible for generating tasks and sampling transmission time, computation time and the vacation duration, strictly according to the given process and distributions.
\end{remark}

\subsection{Experimental analysis of the random delay characteristics} \label{sec-52}

\subsubsection{The variations in the average delay and delay standard deviation with the task transmission rate}

\

In this subsection, we observe the variations in the average delay and delay standard deviation with the task transmission rate under different relationships between $\lambda$ and $\mu_2$. For $\lambda>\mu_2$, the parameters of case 1 in Section \ref{sec-51} are used, except that the task transmission rate varies as in Table \ref{Tab-1}; for $\lambda<\mu_2$, the parameters remain the same as $\lambda>\mu_2$, except that ${{\beta }_{2}}=1$ and ${{S}_{2}}=0.2857$, i.e., ${{\mu }_{2}}=0.7143$ task$/\Delta t$.

\begin{figure}[!t]
\centering
\includegraphics[width=3.4in]{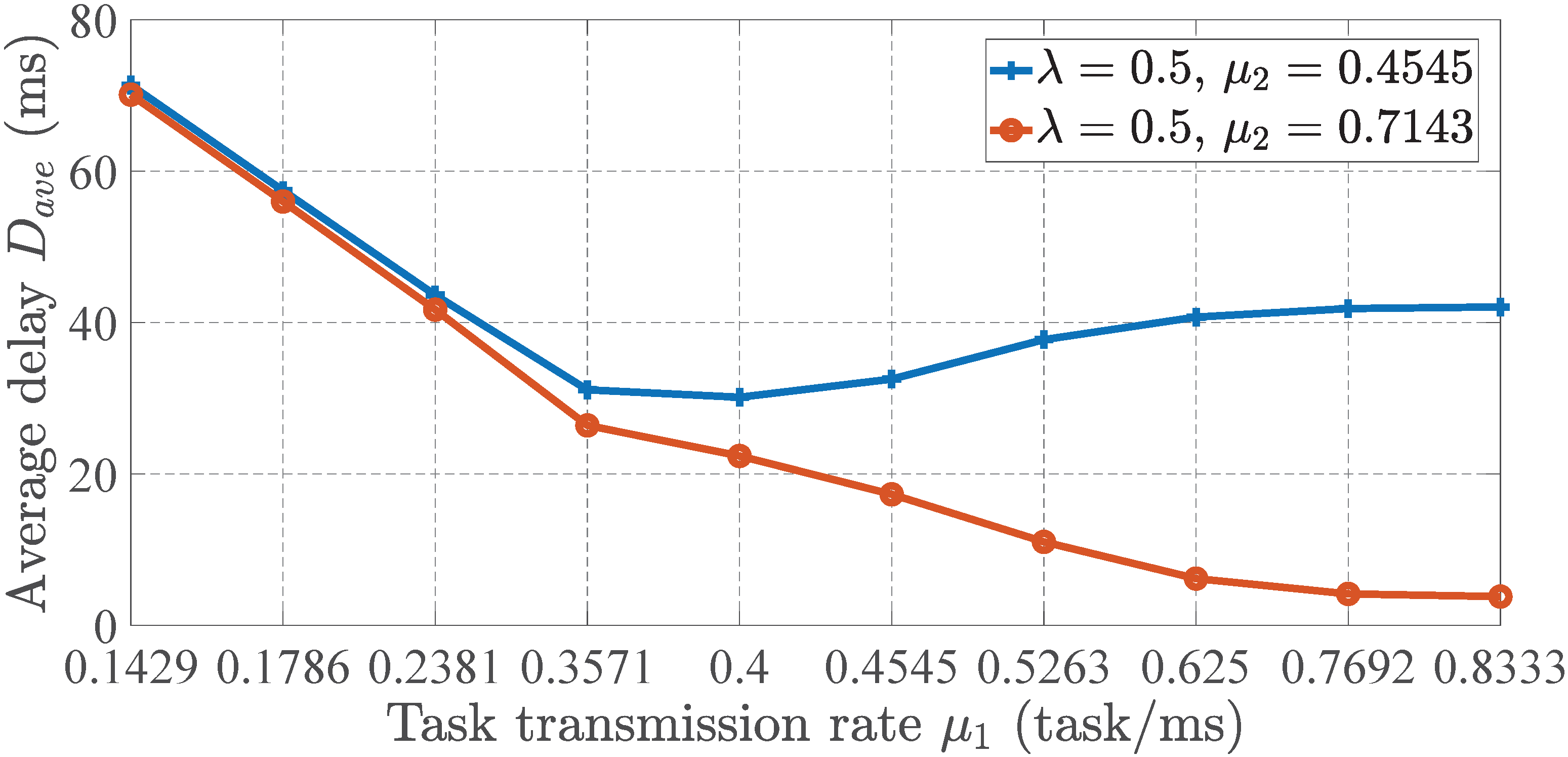}
\caption{The average delay $D_{ave}$ vs. the task transmission rate $\mu_1$ for $\lambda>\mu_2$ and $\lambda<\mu_2$.}
\label{fig-06}
\end{figure}

It is shown in Fig. \ref{fig-06} that for $\lambda>\mu_2$, as the task transmission rate increases, the average delay first decreases, then increases, and finally tends to be constant. For $\lambda<\mu_2$, as the task transmission rate increases, the average delay gradually decreases and then tends to be constant.

\begin{figure}[!t]
\centering
\includegraphics[width=3.4in]{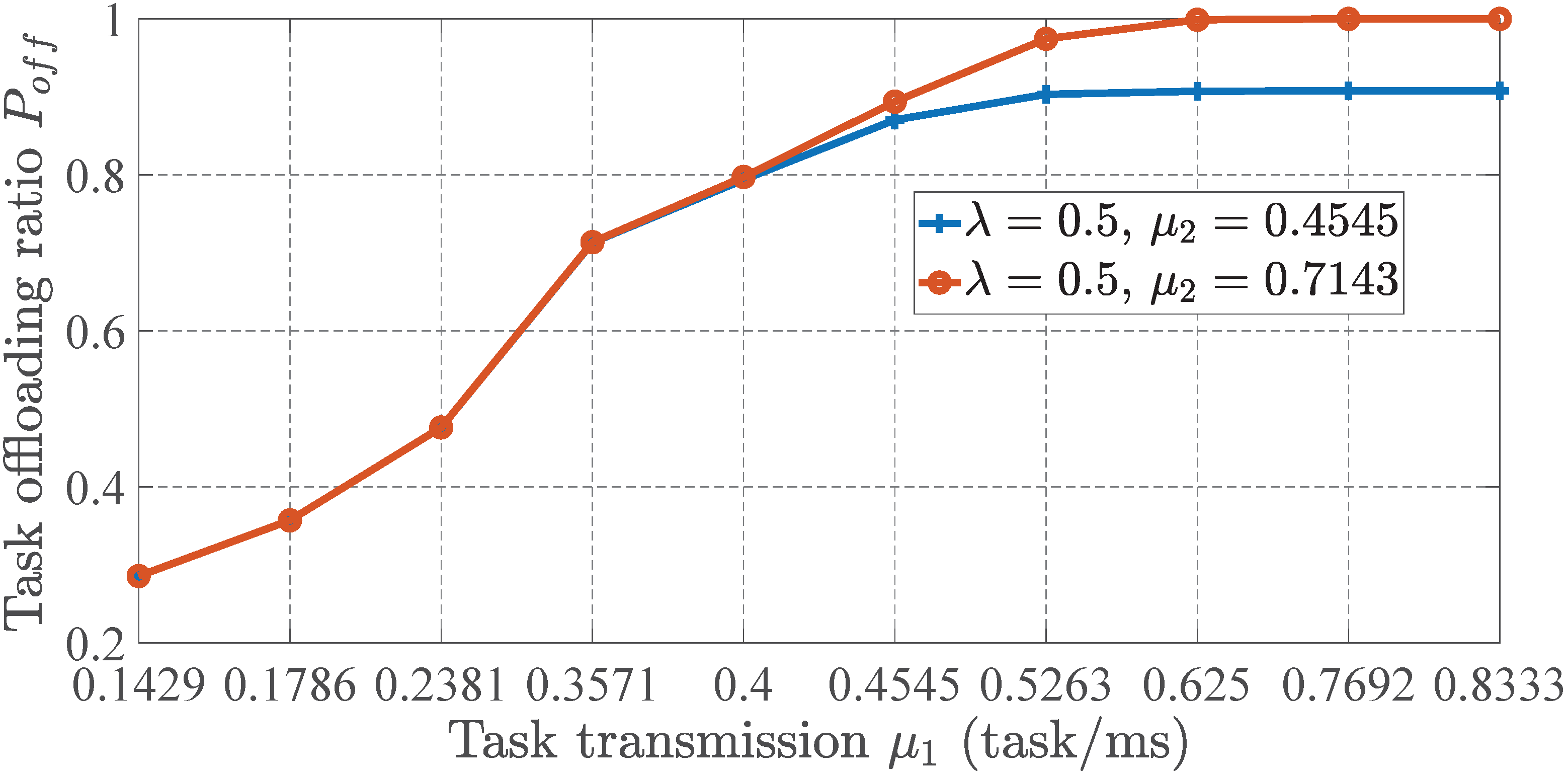}
\caption{The task offloading ratio $P_{off}$ vs. the task transmission rate $\mu_1$ for $\lambda>\mu_2$ and $\lambda<\mu_2$.}
\label{fig-07}
\end{figure}

\begin{figure}[!t]
\centering
\includegraphics[width=3.4in]{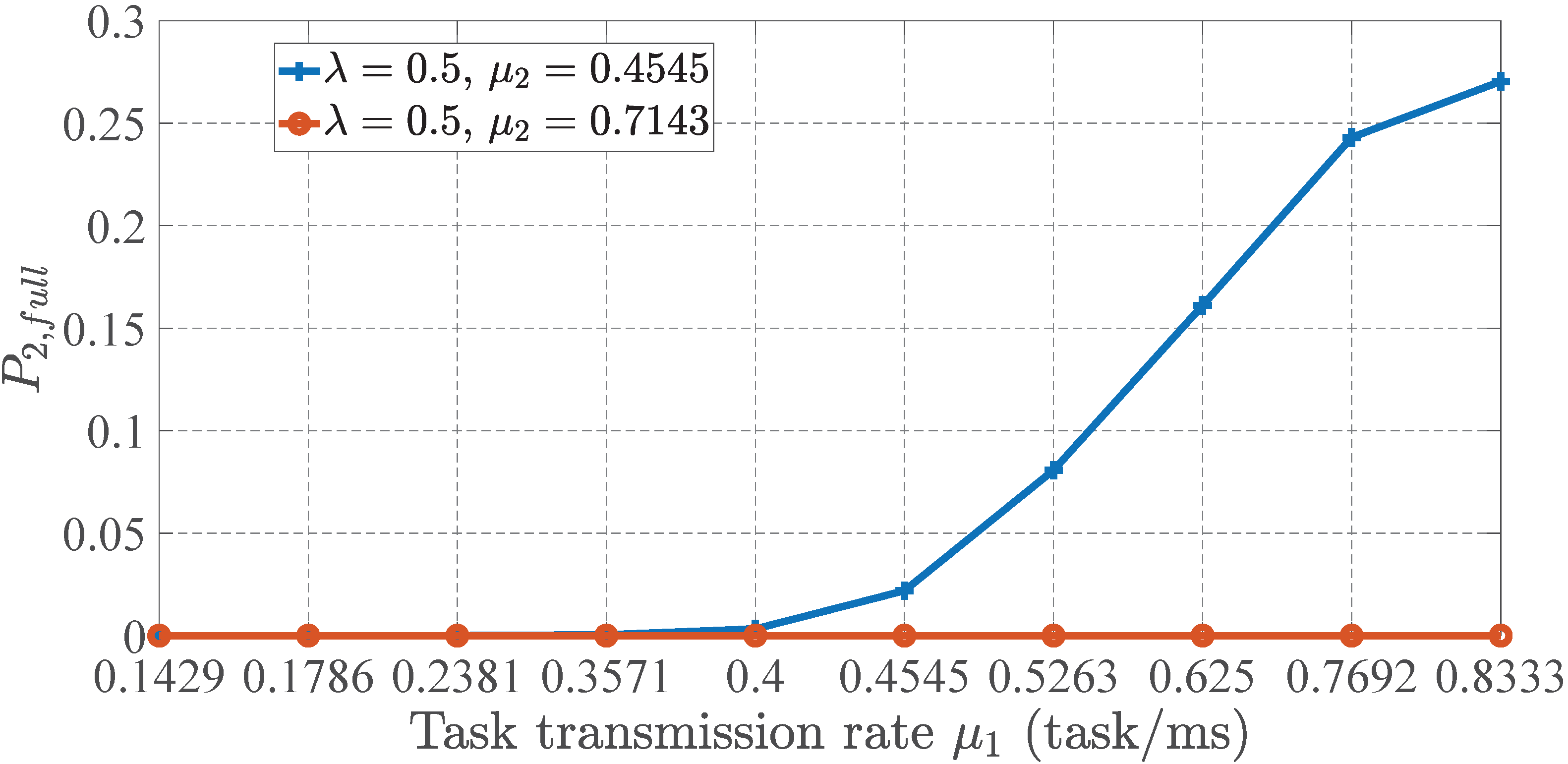}
\caption{The probability $P_{2,full}$ that the computation buffer is full vs. the task transmission rate $\mu_1$ for $\lambda>\mu_2$ and $\lambda<\mu_2$.}
\label{fig-08}
\end{figure}

The reasons for the results shown in Fig. \ref{fig-06} are as follows. The delay consists of four parts, i.e., the transmission waiting time, the transmission time, the computation waiting time and the computation time. As the task transmission rate increases, the transmission time decreases and the computation time remains unchanged; but both the transmission waiting time and the computation waiting time increase, since more tasks can enter the transmission buffer and computation buffer (refer to Fig. \ref{fig-07}, where it is shown that as the task transmission rate increases, the offloading ratio $P_{off}$ increases and then tends to be constant whether $\lambda<\mu_2$ or $\lambda>\mu_2$). If the decrease $\delta^{-}$ of the transmission time is greater than the sum $\delta^{+}$ of the increases of the transmission waiting time and computation waiting time, the average delay decreases; and if the former is less than the latter, the average delay increases. Otherwise, the average delay remains constant. Compared with the case with $\lambda<\mu_2$, in the case with $\lambda>\mu_2$, the computation buffer is prone to saturation (see Fig. \ref{fig-08}), which causes $\delta^{+}=\delta^{-}$ and even $\delta^{+}>\delta^{-}$.

\begin{figure}[!t]
\centering
\includegraphics[width=3.4in]{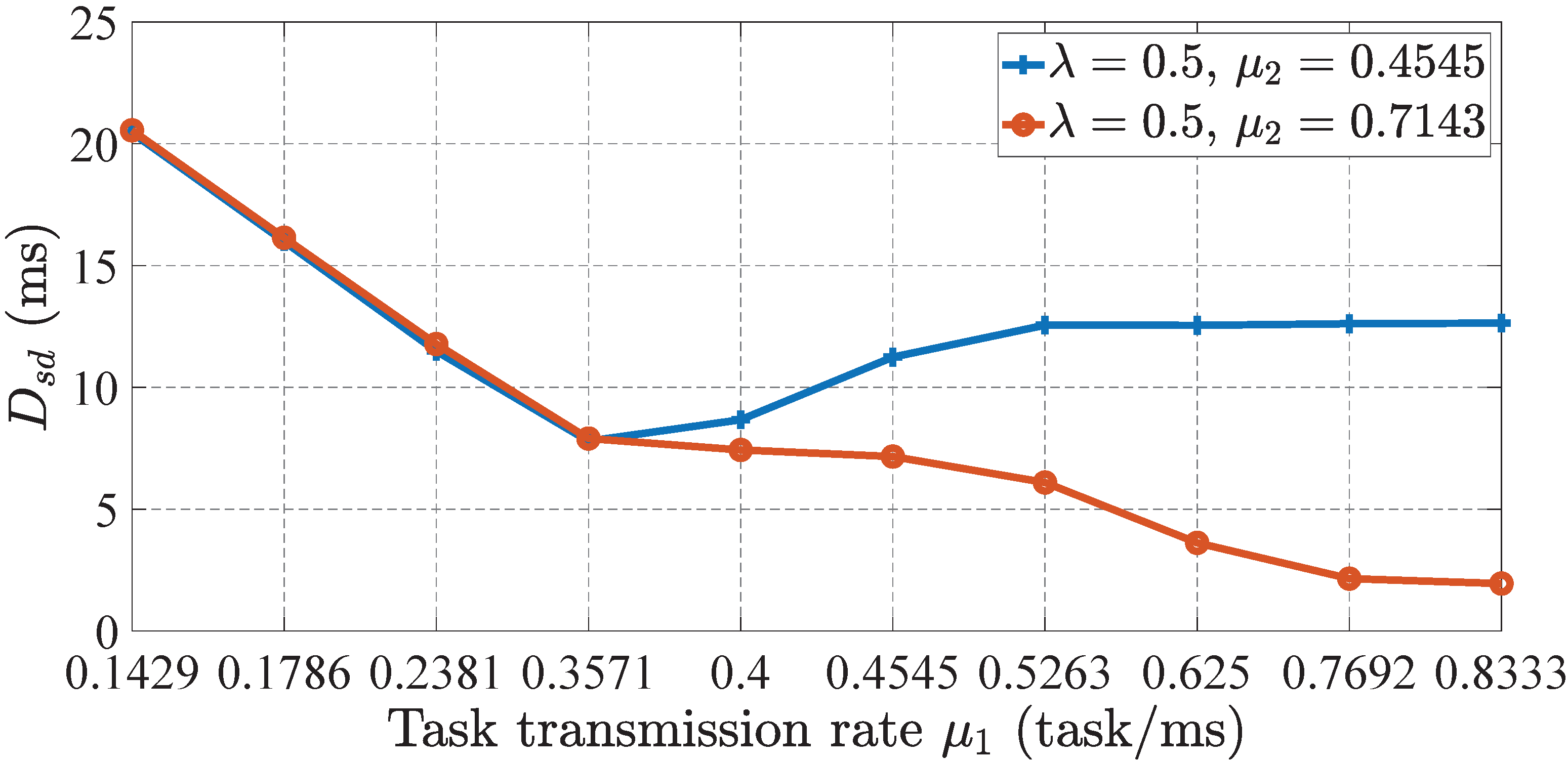}
\caption{The delay standard deviation $D_{sd}$ vs. the task transmission rate $\mu_1$ for $\lambda>\mu_2$ and $\lambda<\mu_2$.}
\label{fig-09}
\end{figure}

It is shown in Fig. \ref{fig-09} that for $\lambda>\mu_2$, as the task transmission rate increases, the delay standard deviation first decreases, then increases, and finally tends to be constant. For $\lambda<\mu_2$, as the task transmission rate increases, the delay standard deviation gradually decreases and then tends to be constant.

The reasons for the results shown in Fig. \ref{fig-09} are as follows. As the task transmission rate increases, more tasks can enter the transmission buffer and computation buffer (see Fig. \ref{fig-07}); meanwhile, the buffering effect of these buffers on the delay fluctuation gradually appears, and so the delay standard deviation decreases. However, the buffering effect will be affected by the computation buffer saturation, which causes the transmission to pause.

It is inferred from the above experiments that for $\lambda>\mu_2$, as the task transmission rate increases, the average delay and delay standard deviation can reach their minimum values when the task transmission rate is close to but does not exceed the computation rate. For $\lambda<\mu_2$, as the task transmission rate increases, the average delay and delay standard deviation gradually decrease and then tend to be constant.

\subsubsection{Characteristics of the probability distribution of the delay}

\

\begin{figure}[!t]
\centering
\includegraphics[width=3.4in]{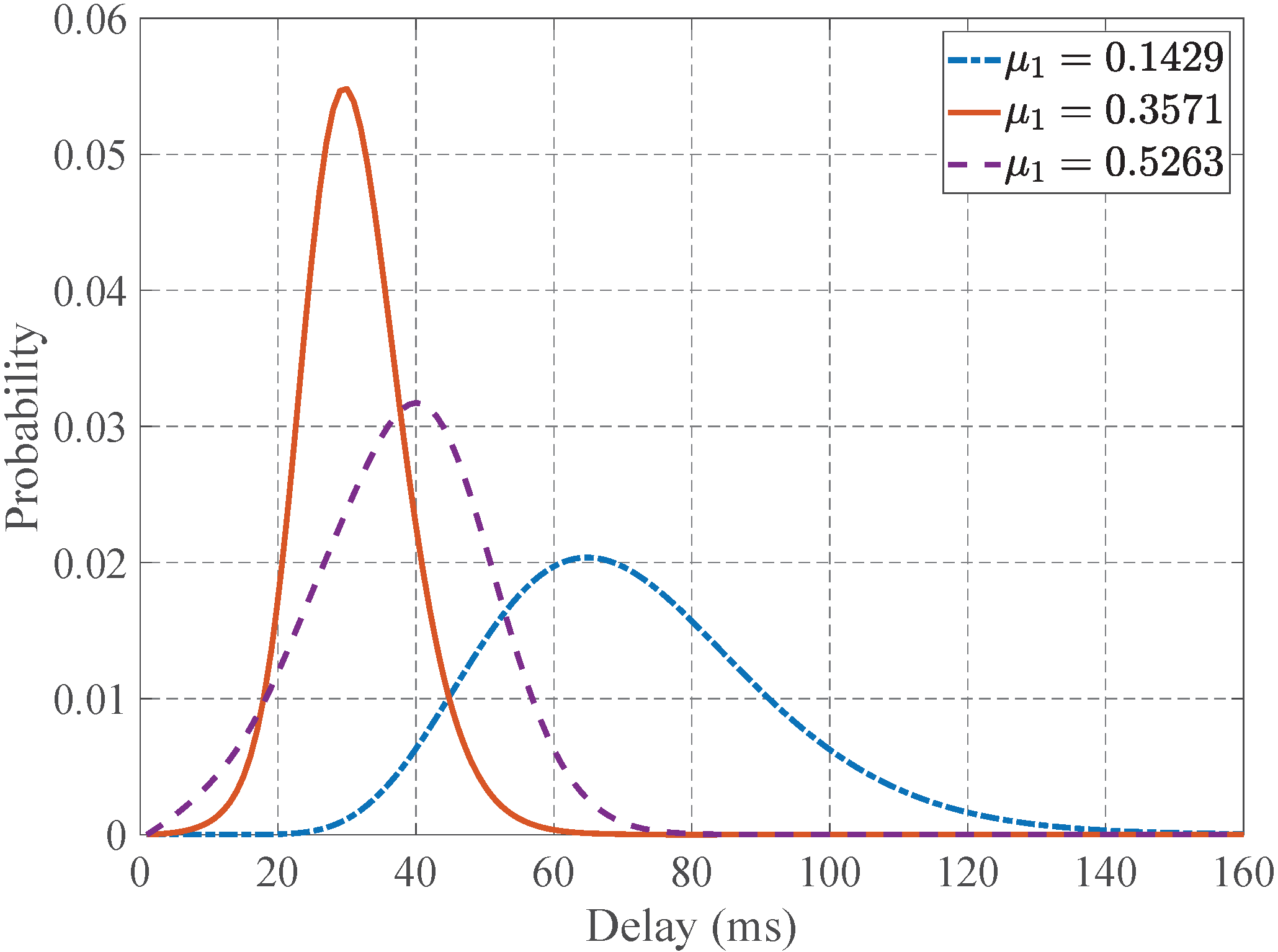}
\caption{The probability distribution of the delay for $\lambda>\mu_2$}
\label{fig-10}
\end{figure}

\begin{figure}[!t]
\centering
\includegraphics[width=3.35in]{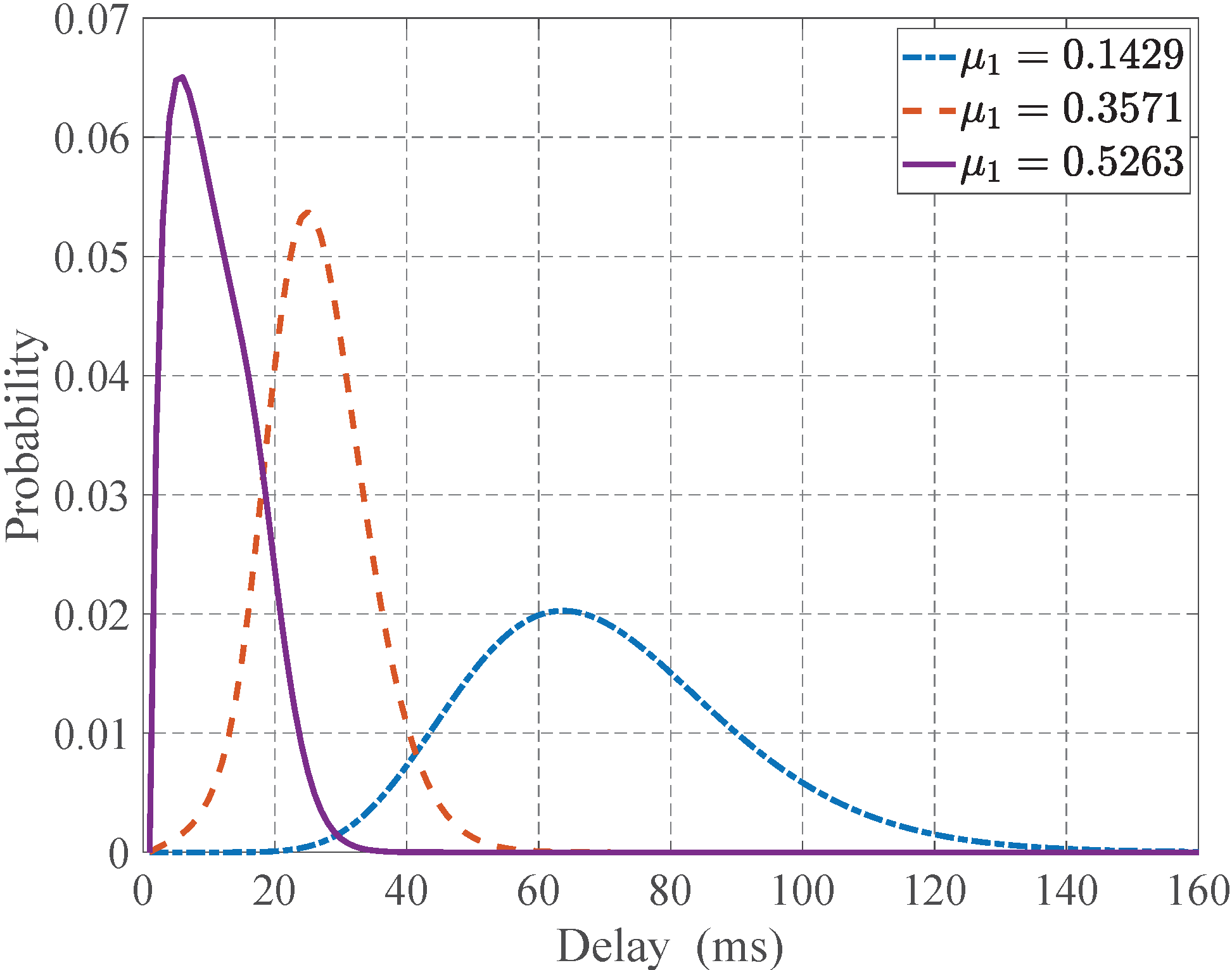}
\caption{The probability distribution of the delay for $\lambda<\mu_2$}
\label{fig-11}
\end{figure}

In this subsection, we observe the characteristics of the probability distribution of the delay under different relationships between $\lambda$ and $\mu_2$. The corresponding parameters in the above subsection are used, except that the task transmission rate varies as $\mu_1=0.1429$, 0.3571 and 0.5263, as shown in Table \ref{Tab-1}. It is shown in Fig. \ref{fig-10} and Fig. \ref{fig-11} that the probability distribution curve of the delay has a single peak and is asymmetric. For $\lambda>\mu_2$, as the task transmission rate increases, the tail of the probability distribution first becomes lighter and then heavier. For $\lambda<\mu_2$, as the task transmission rate increases, the tail of the probability distribution gradually becomes lighter. The reason is similar to the one for delay standard deviation given in the above subsection, since the tail of the probability distribution of delay is also related to the delay fluctuation. For an MEC system, the lighter tail of the probability distribution of the delay and the smaller average delay represent the high-quality TSSs in terms of the delay.


The probability distribution of the delay can comprehensively reflect the delay profile, and can be used to derive all the random delay characteristics. The average delay, the delay standard deviation and the delay bound violation probability are key metrics of the QoS of TSSs in terms of delay. For instance, to ensure the normal operation of industrial automation systems, the delay of industrial control data must not exceed the specified bound, and a low delay jitter (delay standard deviation) is also required. For multimedia streaming systems, the impact of occasional delay bound violations on the quality of multimedia perception can be tolerated. Thus, to guarantee the QoS of some TSSs, random delay characteristics aside from the average delay must be considered in the communication and computation resource allocation. However, if the transmission and computation processes could not be decoupled, the existing estimation methods mostly focused on the average delay. This will degrade the QoS of TSSs. The estimation methods of this paper can be used in optimizing resource allocation for improving QoS compared with only considering the average delay as follows. First, the transmission and computation times are fitted by the D-PH distributions, which are related to the assigned communication and computation resources. Then, according to the estimation methods of this paper, the holistic random delay characteristics are estimated. Finally, the optimization algorithm for the communication and computation resource allocation is established under the constraints of holistic delay requirements.

\section{Discussion} \label{sec-6}

The transition probabilities of the Markov chain form the premise of theoretical analysis in this paper. Their availability depends on the practicability of the queueing model abstracted from the MEC system. However, the more practical the model, the more complex and difficult it is to obtain the transition probabilities. In order to facilitate the theoretical analysis, some assumptions are always made. The analysis results can reflect the delay characteristics of the practical system to a certain extent. Nevertheless, it is necessary to obtain the transition probabilities conforming to the practical system. She \MakeLowercase{\textit{et al.}} \cite{SheCY-2020} merged theoretical models and real data to guarantee the QoS requirement of ultra-reliable and low-latency (URLLC) in 6G networks, by using the model-based analysis results as the guidance and benchmarks for data-driven deep learning. Inspired by this approach, it will be worth trying to combine model-based methods and data-driven methods to analyze the holistic delay characteristics of the MEC system. That is, the transition probabilities obtained through data-driven methods will serve as the inputs of the theoretical analysis of the delay characteristics.

\section{Conclusion and future directions} \label{sec-7}

In this paper, an MEC system with a limited computation buffer at the edge server has been studied. In this system, if the computation buffer is saturated, the UE will pause the task transmission until the computation buffer has new free space. The delay experienced by the task is dominated by the transmission waiting time, the transmission time, the computation waiting time and the computation time. We obtain the estimation methods for the random delay characteristics, including the probability distribution of the delay, the delay bound violation probability, the average delay and the delay standard deviation. These random delay characteristics are key metrics reflecting the QoS of TSSs. Based on the estimation methods given in this paper, when the MEC system allocates the wireless communication bandwidths and CPU clock frequency to the UE running time-sensitive applications, it can consider various delay characteristics in addition to the average delay. This will greatly improve the QoS of TSSs in MEC.

From the system optimization perspective, joint communication and computation resource allocation for guaranteeing the holistic delay performance is an interesting future research direction in practice. Under specified multi-access strategies, the delay estimation model can be established for a more practical multi-user case. The formulated model can also be expanded to the hybrid one with data-driven approaches.

\appendices

\section{The discrete-time Markovian arrival process} \label{Appendix-A}

The D-MAP is governed by a discrete-time Markov chain with the state space $\{1,2,\cdots,m\}$ and the one-step transition probability matrix $D$, where $D=D_0+D_1$ and $D\mathbf{e}=\mathbf{e}$. The matrices $D_0$ and $D_1$ are $m\times m$ nonnegative matrices, and they both have at least one positive entry. For $\forall i,j\in\{1,2,\cdots,m\}$, $(D_0)_{i,j}$ denotes the probability with a transition from state $i$ to $j$ without an arrival; $(D_1)_{i,j}$ denotes the probability with a transition from state $i$ to $j$ with an arrival. The states $1,2,\cdots,m$ are called the phases of the D-MAP. The average arrival rate $\lambda$ is calculated as $\lambda=\bm{\pi}D_{1}\mathbf{e}$, where $\bm{\pi}=\bm{\pi}D$ and $\bm{\pi}\mathbf{e}=1$. More details on D-MAP can be found in \cite{Alfa-2016}.

\section{The PH distribution} \label{Appendix-B}

There are two types of PH distributions: continuous PH (C-PH) distribution and discrete-time PH (D-PH) distribution. They are defined through the continuous Markov chain and the discrete-time Markov chain, respectively.

Consider a continuous Markov chain $\{X(t);t\geq0\}$ with the state space $S_{c}=\{1,2,\cdots,m,m+1\}$ and the infinitesimal generator as
    \begin{equation}
    \mathbf{Q} = \left(
          \begin{array}{cc}
            \mathbf{T} & \mathbf{T}^{0} \\
            \mathbf{0} & 0 \\
          \end{array}
        \right),
    \end{equation}
   where $\mathbf{T}$ is a $m\times m$ matrix and nonsingular, $\mathbf{T}^{0}$ is a $m$-dimensional column vector, and $\mathbf{T}\mathbf{e}+\mathbf{T}^{0}=\mathbf{0}$. The states $1,2,\cdots,m$ are transient states, and the state $m+1$ is the absorbing state. For $\forall i,j\in\{1,2,\cdots,m\}$, $\mathbf{T}_{i,j}$ denotes the transition rate from $i$ to $j$, $\mathbf{T}_{i,i}<0$ and $\mathbf{T}_{i,j}\geq0$ $(i\neq j)$; $\mathbf{T}^{0}_{i}$ denotes the transition rate from $i$ to $m+1$. Given the initial probability vector $(\bm{\alpha},\alpha_{m+1})$ with $\bm{\alpha}\mathbf{e}+\alpha_{m+1}=1$, the probability distribution of time until absorption in the state $m+1$ is called the C-PH distribution with the representation $(\bm{\alpha},\mathbf{T})$ of order $m$. The states $1,2,\cdots,m$ are called the phases of the C-PH distribution. The corresponding distribution function $F(x)$ is given by
   \begin{equation}
     F(x)=1-{\bm{\alpha}}e^{\mathbf{T}x}\mathbf{e}, \quad x\geq0.
   \end{equation}
   The mean ${\zeta}$ is calculated as ${\zeta}=-{\bm{\alpha}}{\mathbf{T}^{-1}}\mathbf{e}$. If ${\bm{\alpha}}=1$ and $\mathbf{T}=-\gamma$, the C-PH distribution is reduced to the exponential distribution with the parameter $\gamma$.

  Similarly, the D-PH distribution is defined by considering a discrete-time Markov chain $\{X(n);n\geq0\}$ with the state space $S_{d}=\{1,2,\cdots,m,m+1\}$ and the one-step transition probability matrix as
    \begin{equation}
    \mathbf{\hat{Q}} = \left(
          \begin{array}{cc}
            \mathbf{\hat{T}} & \mathbf{\hat{T}}^{0} \\
            \mathbf{0} & 1 \\
          \end{array}
        \right),
    \end{equation}
   where $\mathbf{\hat{T}}$ is a $m\times m$ substochastic matrix, such that $\mathbf{I}-\mathbf{\hat{T}}$ is nonsingular, $\mathbf{T}^{0}$ is a $m$-dimensional column vector, and $\mathbf{\hat{T}}\mathbf{e}+\mathbf{\hat{T}}^{0}=\mathbf{e}$. The states $1,2,\cdots,m$ are transient states, and the state $m+1$ is the absorbing state. For $\forall i,j\in\{1,2,\cdots,m\}$, $\mathbf{\hat{T}}_{i,j}$ denotes the transition probability from $i$ to $j$; $\mathbf{T}^{0}_{i}$ denotes the transition probability from $i$ to $m+1$. Given the initial probability vector $(\bm{\alpha},\alpha_{m+1})$, the probability distribution $\{p_{k},k\geq0\}$ of the number $k$ of state transitions until absorption in the state $m+1$ is called the D-PH distribution with the representation $(\bm{\alpha},\mathbf{\hat{T}})$ of order $m$. The states $1,2,\cdots,m$ are called the phases of the D-PH distribution. The probability distribution $\{p_{k},k\geq0\}$ of the D-PH distribution is given by
   \begin{align}
     & p_{0} = \alpha_{m+1},\\
     & p_{k} = \bm{\alpha}\mathbf{\hat{T}}^{k-1}\mathbf{\hat{T}}^{0}, \quad k\geq1.
   \end{align}
The mean $\hat{\zeta}$ is calculated as $\hat{\zeta}=\bm{\alpha}(\mathbf{I}-\mathbf{\hat{T}})^{-1}\mathbf{e}$.

It is shown in Theorem 9.14 of \cite{BreuerBaum-2005} that the class of PH distributions is dense within the class of all distributions on $[0,+\infty)$. More details on PH distribution can be found in \cite{Neuts-73,BreuerBaum-2005}.

\section{The method for obtaining the stationary distribution of the Markov chain $\Xi$} \label{Appendix-C}

The system of equations \eqref{eq-StaDis} can be rewritten as follows,
\begin{align}
    & {{\mathbf{x}}_{0}}={{\mathbf{x}}_{0}}{{M}_{0}}+{{\mathbf{x}}_{1}}{{M}_{2}}, \label{eq-StaDis-S1}\\
    & {{\mathbf{x}}_{1}}={{\mathbf{x}}_{0}}{{M}_{1}}+{{\mathbf{x}}_{1}}{{M}_{3}}+{{\mathbf{x}}_{2}}{{M}_{5}}, \label{eq-StaDis-S2}\\
    & {{\mathbf{x}}_{i_1}}={{\mathbf{x}}_{i_1-1}}{{M}_{4}}+{{\mathbf{x}}_{i_1}}{{M}_{3}}+{{\mathbf{x}}_{i_1+1}}{{M}_{5}}, \nonumber\\
    &  \phantom{{{\mathbf{x}}_{i_1}}={{\mathbf{x}}_{i_1-1}}{{M}_{4}}+{{\mathbf{x}}_{i_1}}{{M}_{3}}} 2\le i_1\le {{N}_{1}}-1,  \label{eq-StaDis-S3}\\
    & {{\mathbf{x}}_{{{N}_{1}}}}={{\mathbf{x}}_{{{N}_{1}}-1}}{{M}_{4}}+{{\mathbf{x}}_{{{N}_{1}}}}{M'_{3}}, \label{eq-StaDis-S4} \\
    & \sum\limits_{{{i}_{1}}=0}^{{{N}_{1}}}{{{\mathbf{x}}_{{{i}_{1}}}}\mathbf{e}}=1. \label{eq-StaDis-S4a}
\end{align}
Based on the matrix-geometric method \cite{Neuts-73}, there exists a matrix $R$ such that
\begin{equation}
  R={{M}_{4}}+R{{M}_{3}}+{{R}^{2}}{{M}_{5}}.
\end{equation}
$R$ can be obtained as the following successive substitution procedure,
\begin{align}
  & R_{(0)}=\mathbf{0}, \\
  & R_{(n+1)}={{M}_{4}}+R_{(n)}{{M}_{3}}+{R^{2}_{(n)}}{{M}_{5}}, \quad n=0,1,\cdots,
\end{align}
until the maximum entry-wise difference between two consecutive approximations is less than an allowed error, where $n$ is the index of iteration. From equations \eqref{eq-StaDis-S3} and \eqref{eq-StaDis-S4}, there are the following relations,
\begin{align}
  & {{\mathbf{x}}_{{{i}_{1}}}}={{\mathbf{x}}_{{{i}_{1}}-1}}R, \quad 2\le i_1\le {{N}_{1}}-1, \label{eq-StaDis-S5} \\
  & {{\mathbf{x}}_{{{N}_{1}}}}={{\mathbf{x}}_{{{N}_{1}}-1}}{{M}_{4}}{{\left( \mathbf{I}-{M'_{3}} \right)}^{-1}}. \label{eq-StaDis-S6}
\end{align}
From equations \eqref{eq-StaDis-S1}, \eqref{eq-StaDis-S2}, \eqref{eq-StaDis-S4a}, \eqref{eq-StaDis-S5} and \eqref{eq-StaDis-S6}, ${{\mathbf{x}}_{0}}$ and ${{\mathbf{x}}_{1}}$ satisfy
\begin{align}
  & \left( \begin{matrix}
    {{\mathbf{x}}_{0}} & {{\mathbf{x}}_{1}}  \\
    \end{matrix} \right)\left( \begin{matrix}
    {{M}_{0}} & {{M}_{1}}  \\
    {{M}_{2}} & {{M}_{3}}+R{{M}_{\text{5}}}  \\
    \end{matrix} \right)=\left( \begin{matrix}
    {{\mathbf{x}}_{0}} & {{\mathbf{x}}_{1}}  \\
    \end{matrix} \right), \\
  & \;\,{{\mathbf{x}}_{0}}\mathbf{e}+{{\mathbf{x}}_{1}}\left[ \sum\limits_{n=0}^{{{N}_{1}}-2}{{{R}^{n}}}+{{R}^{{{N}_{1}}-2}}{{M}_{4}}{{\left( \mathbf{I}-{M'_{3}} \right)}^{-1}} \right]\mathbf{e}=1.
\end{align}
${{\mathbf{x}}_{0}}$ and ${{\mathbf{x}}_{1}}$ can be obtained by solving the above system of equations. When the dimension size is large, iterative methods are usually more efficient and more effective than the direct methods \cite{Alfa-2016}. Based on the Jacobi method, ${{\mathbf{x}}_{0}}$ and ${{\mathbf{x}}_{1}}$ can be obtained as the following iterative process.

First, separate ${{\left( {{M}_{0}}-\mathbf{I} \right)}^{T}}$ into a diagonal component $\hat{D}_{0}$, an upper triangular component $\hat{U}_{0}$ and a lower triangular component $\hat{L}_{0}$, respectively, i.e., ${{\left( {{M}_{0}}-\mathbf{I} \right)}^{T}}=\hat{D}_{0}+\hat{U}_{0}+\hat{L}_{0}$. Second, separate $\left({{M}_{3}}+R{{M}_{\text{5}}}-\mathbf{I}\right)^{T}$ into a diagonal component $\hat{D}_{3}$, an upper triangular component $\hat{U}_{3}$ and a lower triangular component $\hat{L}_{3}$, respectively, i.e., $\left({{M}_{3}}+R{{M}_{\text{5}}}-\mathbf{I}\right)^{T}=\hat{D}_{3}+\hat{U}_{3}+\hat{L}_{3}$. Third, apply the following relationships until the maximum entry-wise difference between two consecutive approximations is less than an allowed error.
\begin{equation}
  {{\left( \begin{matrix}
    \mathbf{\hat{x}}_{0}^{(n+1)} & \mathbf{\hat{x}}_{1}^{(n+1)}  \\
    \end{matrix} \right)}^{T}}=-{{\hat{D}}^{-1}}\left( \hat{L}+\hat{U}\right){{\left( \begin{matrix}
    \mathbf{\hat{x}}_{0}^{(n)} & \mathbf{\hat{x}}_{1}^{(n)}  \\
    \end{matrix} \right)}^{T}},
\end{equation}
where $n$ is the index of iteration, $n=0,1,\cdots$; $\mathbf{\hat{x}}_{0}^{(0)}\mathbf{e}+\mathbf{\hat{x}}_{1}^{(0)}\mathbf{e}=1$;
\begin{equation*}
    \hat{D}=\left( \begin{matrix}
        {{\hat{D}}_{0}} & \mathbf{0}  \\
        \mathbf{0} & {{\hat{D}}_{3}}  \\
      \end{matrix} \right), \hat{L}=\left( \begin{matrix}
       {{\hat{L}}_{1}} & \mathbf{0}  \\
        M_{1}^{T} & {{\hat{L}}_{3}}  \\
     \end{matrix} \right), \hat{U}=\left( \begin{matrix}
      {{\hat{U}}_{1}} & M_{2}^{T}  \\
      \mathbf{0} & {{\hat{U}}_{3}}  \\
     \end{matrix} \right).
\end{equation*}
Finally, if the above iteration stops with $n=N$, then
\begin{equation}
  \left( \begin{matrix}
    \mathbf{{x}}_{0} & \mathbf{{x}}_{1}  \\
    \end{matrix} \right)=\left( \begin{matrix}
    \mathbf{\hat{x}}_{0}^{(N+1)} & \mathbf{\hat{x}}_{1}^{(N+1)}  \\
    \end{matrix} \right)/\sigma,
\end{equation}
where
\begin{align}
  \sigma =& \mathbf{\hat{x}}_{0}^{(N+1)}\mathbf{e} +\mathbf{\hat{x}}_{1}^{(N+1)} \nonumber\\
  &\times\left[ \sum\limits_{n=0}^{{{N}_{1}}-2}{{{R}^{n}}}+{{R}^{{{N}_{1}}-2}}{{M}_{4}}{{\left( \mathbf{I}-{M'_{3}} \right)}^{-1}} \right]\mathbf{e}.
\end{align}

The iterative process of the abvoe Jacobi method can be executed in parallel.

\section{The detailed expression of each matrix block in matrix $P$} \label{Appendix-D}

  For matrix ${M}_{0}$,
    \begin{align*}
        & {{M}_{00}}={{D}_{0}}, \\
        & {{M}_{01}}={{D}_{0}}\otimes S_{2}^{0},\\
        & {{M}_{02}}={{D}_{0}}\otimes {{S}_{2}},\\
        & {{M}_{03}}={{D}_{0}}\otimes \left( S_{2}^{0}{{\beta }_{2}} \right),\\
        & M_{02}^{*}={{M}_{02}},
    \end{align*}
  where $S_{2}^{0}=\mathbf{e}-S_{2}\mathbf{e}$.

  For matrix ${M}_{1}$,
    \begin{align*}
        & {{M}_{10}}=\left( \begin{matrix}
          \mathbf{0}_{m\times\tau_3} & {{D}_{1}}\otimes {{\beta }_{1}}  \\
          \end{matrix} \right),\\
        & {{M}_{11}}=\left( \begin{matrix}
          \mathbf{0}_{\tau_4\times\tau_3} & {{D}_{1}}\otimes {{\beta }_{1}}\otimes S_{2}^{0} \\
          \end{matrix} \right),\\
        & {{M}_{12}}=\left( \begin{matrix}
          \mathbf{0}_{\tau_4\times\tau_2} & {{D}_{1}}\otimes {{\beta }_{1}}\otimes S_{2} \\
          \end{matrix} \right),\\
        & {{M}_{13}}=\left( \begin{matrix}
          \mathbf{0}_{\tau_4\times\tau_2} & {{D}_{1}}\otimes {{\beta }_{1}}\otimes \left( S_{2}^{0}{{\beta }_{2}}\right) \\
          \end{matrix} \right),\\
        & M_{12}^{*}={{D}_{1}}\otimes v\otimes {{S}_{2}},
    \end{align*}
    where $\tau_3 = m\cdot {{l}_{2}}$ and $\tau_4=m\cdot n_2$.

  For matrix ${M}_{2}$,
    \begin{align*}
        & {{M}_{20}}=\begin{pmatrix}
          {{\mathbf{0}}_{\tau_3\times\tau_4}}  \\
              \begin{array}{l}
                {{D}_{0}}\otimes \left( S_{1}^{0}{{\beta }_{2}} \right)
              \end{array}
          \end{pmatrix},\\
        & {{M}_{21}}=\begin{pmatrix}
          {{\mathbf{0}}_{\tau_2\times\tau_4}}  \\
          {{D}_{0}}\otimes S_{1}^{0}\otimes \left( S_{2}^{0}{{\beta }_{2}} \right)
          \end{pmatrix},\\
        & {{M}_{22}}=\begin{pmatrix}
          {{\mathbf{0}}_{\tau_2\times\tau_4}} \\
          {{D}_{0}}\otimes S_{1}^{0}\otimes {{S}_{2}}
          \end{pmatrix},\\
        & M_{21}^{*}={{\mathbf{0}}_{\tau_2\times\tau_4}}.
    \end{align*}

   For matrix ${M}_{3}$,
    \begin{align*}
        & {{M}_{30}}=\left( \begin{matrix}
          {{D}_{0}}\otimes V & {{D}_{0}}\otimes \left( {{V}^{0}}\beta _{1}  \right)\\
          \mathbf{0} & {{D}_{0}}\otimes {{S}_{1}}  \\
          \end{matrix} \right),\\
        & {{M}_{31}}=\begin{pmatrix}
          {{\mathbf{0}}_{\tau_3\times\tau_2}} & {{\mathbf{0}}_{\tau_3\times\tau_5}}  \\
          \mathbf{0} & {{D}_{1}}\otimes \left( S_{1}^{0}{{\beta }_{1}} \right)\otimes {{\beta }_{2}}
          \end{pmatrix},\\
        & {{M}_{32}}=\begin{pmatrix}
          {{D}_{0}}\otimes V\otimes S_{2}^{0} & {{D}_{0}}\otimes {{V}^{0}}\otimes \left( S_{2}^{0}\beta _{1}\right)\\
          \mathbf{0}&{{D}_{0}}\otimes {{S}_{1}}\otimes S_{2}^{0}
          \end{pmatrix},\\
        & {{M}_{33}}=\begin{pmatrix}
           {{D}_{0}}\otimes V\otimes {{S}_{2}} &{{D}_{0}}\otimes {{V}^{0}}\otimes \beta _{1}\otimes{{S}_{2}}\\
            \mathbf{0}& \begin{array}{l}
                {{D}_{0}}\otimes {{S}_{1}} \otimes {{S}_{2}}\\
                \;\,+{{D}_{1}}\otimes \left( S_{1}^{0}{{\beta }_{1}} \right)\\
                \;\,\otimes \left( S_{2}^{0}{{\beta }_{2}} \right)
              \end{array}
          \end{pmatrix},\\
        & {{M}_{34}}=\begin{pmatrix}
           {{\mathbf{0}}_{\tau_2\times\tau_2}} & {{\mathbf{0}}_{\tau_2\times\tau_5}}  \\
            \mathbf{0}
            & {D}_{1}\otimes \left( S_{1}^{0}{{\beta }_{1}} \right)\otimes {{S}_{2}}
          \end{pmatrix},\\
        & {{M}_{35}}=\begin{pmatrix}
            \begin{array}{l}
              {{D}_{0}}\otimes V \\
              \;\;\otimes \left( S_{2}^{0}{{\beta }_{2}} \right)
            \end{array}
            & \begin{array}{l}
                {{D}_{0}}\otimes {{V}^{0}}\otimes \beta _{1} \\
                \;\;\otimes \left( S_{2}^{0}{{\beta }_{2}} \right)
              \end{array}\\
            \mathbf{0}
            & {{D}_{0}}\otimes {{S}_{1}}\otimes \left( S_{2}^{0}{{\beta }_{2}} \right)
          \end{pmatrix},\\
        & M_{34}^{*}=\begin{pmatrix}
           {{\mathbf{0}}_{\tau_2\times\tau_2}}  \\
             {{D}_{1}}\otimes v \otimes S_{1}^{0}\otimes {{S}_{2}}
          \end{pmatrix},\\
        & M_{35}^{*}=\begin{pmatrix}
           \begin{array}{l}
             {{D}_{0}}\otimes V \\
             \;\,\otimes \left( S_{2}^{0}{{\beta }_{2}} \right)
           \end{array}
            & \begin{array}{l}
            {{D}_{0}}\otimes {{V}^{0}}\otimes \beta _{1}\\
            \;\,\otimes \left( S_{2}^{0}{{\beta }_{2}} \right)
            \end{array}
          \end{pmatrix},\\
        & M_{33}^{*}={{D}_{0}}\otimes V\otimes {{S}_{2}}+{{D}_{0}}\otimes \left( {{V}^{0}}v \right)\otimes {{S}_{2}},
    \end{align*}
    where $\tau_5 = m\cdot {{n}_{1}}\cdot {{n}_{2}}$ and $S_{1}^{0}=\mathbf{e}-S_{1}\mathbf{e}$.

   For matrix ${M'_{3}}$,
      \begin{align*}
        & {M'_{30}}=\begin{pmatrix}
           D\otimes V
             & D\otimes\left({{V}^{0}}\beta _{1}\right)\\
           \mathbf{0} & D\otimes {{S}_{1}}
         \end{pmatrix},\\
        & {M'_{31}}=\begin{pmatrix}
           {{\mathbf{0}}_{\tau_3\times\tau_2}} & {{\mathbf{0}}_{\tau_3\times\tau_5}}  \\
            \mathbf{0}
            & {{D}_{1}}\otimes \left( S_{1}^{0}{{\beta }_{1}} \right)\otimes {{\beta }_{2}}
          \end{pmatrix},\\
        & {M'_{32}}=\begin{pmatrix}
            D \otimes V\otimes S_{2}^{0} & D \otimes {{V}^{0}}\otimes \left( S_{2}^{0}\beta _{1} \right)\\
           \mathbf{0}&D\otimes {{S}_{1}}\otimes S_{2}^{0}
          \end{pmatrix},\\
        & {M'_{33}}=\begin{pmatrix}
            D \otimes V\otimes {{S}_{2}}&D\otimes {{V}^{0}}\otimes \beta _{1}\otimes {{S}_{2}}\\
           \mathbf{0}
           & \begin{array}{l}
              D\otimes {{S}_{1}}\otimes {{S}_{2}} \\
              \;\,+{{D}_{1}}\otimes \left( S_{1}^{0}{{\beta }_{1}} \right)\\
              \;\,\otimes \left( S_{2}^{0}{{\beta }_{2}} \right)
             \end{array}
        \end{pmatrix},\\
      & {M'_{34}}=\begin{pmatrix}
         {{\mathbf{0}}_{\tau_2\times\tau_2}} & {{\mathbf{0}}_{\tau_2\times\tau_5}}  \\
          \mathbf{0}& {{D}_{1}}\otimes \left( S_{1}^{0}{{\beta }_{1}} \right)\otimes {{S}_{2}}
        \end{pmatrix}\\
      & {M'_{35}}=\begin{pmatrix}
          \begin{array}{l}
            D\otimes V \\
            \;\,\otimes \left( S_{2}^{0}{{\beta }_{2}} \right)
          \end{array}
      & \begin{array}{l}
          D\otimes {{V}^{0}} \otimes \beta _{1}\\
          \;\,\otimes \left( S_{2}^{0}{{\beta }_{2}} \right)
        \end{array}\\
        \mathbf{0}
      & D\otimes {{S}_{1}}\otimes \left( S_{2}^{0}{{\beta }_{2}} \right)
       \end{pmatrix},\\
      & M'^{*}_{34}=\begin{pmatrix}
         {{\mathbf{0}}_{\tau_2\times\tau_2}}  \\
         {{D}_{1}}\otimes v\otimes S_{1}^{0}\otimes {{S}_{2}}
        \end{pmatrix},\\
      & M'^{*}_{35}=\begin{pmatrix}
          \begin{array}{l}
            D\otimes V\\
            \;\,\otimes \left( S_{2}^{0}{{\beta }_{2}} \right)
          \end{array}
        & \begin{array}{l}
            D\otimes {{V}^{0}}\otimes \beta _{1} \\
            \;\,\otimes \left( S_{2}^{0}{{\beta }_{2}} \right)
          \end{array}
        \end{pmatrix},\\
      & M'^{*}_{33}=D\otimes V\otimes {{S}_{2}}+D\otimes \left( {{V}^{0}}v \right)\otimes {{S}_{2}}.
      \end{align*}

   For matrix ${M_{4}}$,
      \begin{align*}
        & {{M}_{40}}=\begin{pmatrix}
           {{D}_{1}}\otimes V & {{D}_{1}}\otimes {{V}^{0}}\otimes \beta _{1}\\
           \mathbf{0}& {{D}_{1}}\otimes {{S}_{1}}  \\
          \end{pmatrix},\\
        & {{M}_{41}}=\begin{pmatrix}
           {{D}_{1}}\otimes V\otimes S_{2}^{0}
           & {{D}_{1}}\otimes{{V}^{0}}\otimes \left( S_{2}^{0}\beta _{1} \right)\\
            \mathbf{0}& {{D}_{1}}\otimes {{S}_{1}}\otimes S_{2}^{0}
          \end{pmatrix},\\
        & {{M}_{42}}=\begin{pmatrix}
           {{D}_{1}}\otimes V\otimes {{S}_{2}}
           & {{D}_{1}}\otimes{{V}^{0}}\otimes \beta _{1}\otimes {{S}_{2}}\\
             \mathbf{0}& {{D}_{1}}\otimes {{S}_{1}}\otimes {{S}_{2}}
          \end{pmatrix},\\
        & {{M}_{43}}=\begin{pmatrix}
             \begin{array}{l}
               {{D}_{1}}\otimes V  \\
               \;\,\otimes \left( S_{2}^{0}{{\beta }_{2}} \right)
             \end{array}
           & \begin{array}{l}
               {{D}_{1}}\otimes {{V}^{0}}\otimes \beta _{1} \\
               \;\,\otimes \left( S_{2}^{0}{{\beta }_{2}} \right)
             \end{array} \\
            \mathbf{0}
           & {{D}_{1}}\otimes {{S}_{1}} \otimes \left( S_{2}^{0}{{\beta }_{2}} \right)
          \end{pmatrix},\\
        & M_{43}^{*}=\begin{pmatrix}
           \begin{array}{l}
             {{D}_{1}}\otimes V \\
             \;\,\otimes \left( S_{2}^{0}{{\beta }_{2}} \right)
           \end{array}
           & \begin{array}{l}
              {{D}_{1}}\otimes {{V}^{0}}\otimes \beta _{1} \\
              \;\,\otimes \left( S_{2}^{0}{{\beta }_{2}} \right)
             \end{array}
          \end{pmatrix},\\
        & M_{42}^{*}={{D}_{1}}\otimes V\otimes {{S}_{2}}+{{D}_{1}}\otimes \left( {{V}^{0}}v \right)\otimes {{S}_{2}}.
      \end{align*}

   For matrix ${M_{5}}$,
    \begin{align*}
      & {{M}_{50}}=\begin{pmatrix}
         {{\mathbf{0}}_{\tau_2\times\tau_2}} & {{\mathbf{0}}_{\tau_2\times\tau_5}}  \\
          \mathbf{0}
          & {{D}_{0}}\otimes \left( S_{1}^{0}{{\beta }_{1}} \right)\otimes {{\beta }_{2}}
        \end{pmatrix},\\
      & {{M}_{51}}=\begin{pmatrix}
           {{\mathbf{0}}_{\tau_2\times\tau_2}} & {{\mathbf{0}}_{\tau_2\times\tau_5}}  \\
            \mathbf{0}&{{D}_{0}}\otimes \left( S_{1}^{0}{{\beta }_{1}} \right)\otimes \left( S_{2}^{0}{{\beta }_{2}} \right)
        \end{pmatrix},\\
      & {{M}_{52}}=\begin{pmatrix}
           {{\mathbf{0}}_{\tau_2\times\tau_2}} & {{\mathbf{0}}_{\tau_2\times\tau_5}}  \\
            {\mathbf{0}}
            &{{D}_{0}}\otimes \left( S_{1}^{0}{{\beta }_{1}} \right)\otimes {{S}_{2}}
        \end{pmatrix},\\
      & M_{52}^{*}=\begin{pmatrix}
           {{\mathbf{0}}_{\tau_2\times\tau_2}}  \\
           {{D}_{0}}\otimes v\otimes S_{1}^{0}\otimes {{S}_{2}}
        \end{pmatrix},\\
      & M_{51}^{*}={{\mathbf{0}}_{\tau_2\times\tau_2}}.
    \end{align*}

\bibliographystyle{IEEEtran}
\bibliography{IEEEabrv,MEC_Ref}

\end{document}